\documentclass[aps,pra,twocolumn,nofootinbib,showpacs]{revtex4}
\usepackage{amsbsy,latexsym}
\usepackage{amsfonts}
\usepackage{amssymb}
\usepackage[mathscr]{eucal}
\usepackage{epsfig,graphics,graphicx}
\newcommand{\ketbrad}[1]{|#1\rangle\!\langle #1|}

\newcommand{\al}{\alpha}
\newcommand{\be}{\begin{equation}}
\newcommand{\ee}{\end{equation}}

\newcommand{\half}{\frac{1}{2}}

\newcommand{\ex}[1]{\mathrm{e}^{#1}}
\newcommand{\id}{\openone}

\newcommand{\tr}{{\rm tr}}

\newcommand{\ket}[1]{\vert #1 \rangle}
\newcommand{\bra}[1]{\langle #1 \vert}

\newcommand{\thb}{{\boldsymbol \theta}}

\newcommand{\ea}{\emph{et al.}}
\newcommand{\gb}{{\boldsymbol g}}

\newtheorem{theorem}{Theorem}
\newtheorem{lemma}{Lemma}
\newcommand{\eqref}[1]{(\ref{#1})}

\newcommand{\pcher}{P_{\rm QC}}

\begin{document}

\title{The quantum Chernoff bound as a measure of distinguishability between
density matrices: application to qubit and Gaussian states}

\author{J.~Calsamiglia}
\affiliation{Grup de F\'{\i}sica Te\`{o}rica, Universitat Aut\`{o}noma de
Barcelona, 08193 Bellaterra (Barcelona), Spain}
\author{Ll.~Masanes}
\affiliation{Department of Applied Mathematics and Theoretical
Physics, University of Cambridge, Wilberforce Road, Cambridge CB3
0WA, U.K.}
\author{R.~Mu\~{n}oz-Tapia}
\affiliation{Grup de F\'{\i}sica Te\`{o}rica, Universitat Aut\`{o}noma de
Barcelona, 08193 Bellaterra (Barcelona), Spain}

\author{A.~Acin}
\affiliation{ICREA and ICFO-Institut de Ciencies Fotoniques, Mediterranean
Technology Park, 08860 Castelldefels (Barcelona), Spain}

\author{E.~Bagan}
\affiliation{Department  of Physics and Astronomy, University of New
Mexico, Albuquerque , NM 87131, USA} \affiliation{Grup de F\'{\i}sica
Te\`{o}rica, Universitat Aut\`{o}noma de Barcelona, 08193 Bellaterra
(Barcelona), Spain}

\date{\today}

\begin{abstract}
Hypothesis testing is a fundamental issue in statistical inference
and has been a crucial element in the development of information
sciences. The Chernoff bound gives the minimal Bayesian error
probability when discriminating two hypotheses given a large number
of observations. Recently the combined work of Audenaert \ea~[Phys.
Rev. Lett. {\bf 98}, 160501]  and Nussbaum and
Szkola~[quant-ph/0607216] has proved the quantum analog of this
bound, which applies  when the hypotheses correspond to two quantum
states.  Based on the quantum Chernoff bound, we define a physically meaningful
distinguishability measure and its corresponding metric in the space
of states; the latter is shown to coincide with the Wigner-Yanase metric. 
Along the same lines, we define a second, more easily implementable, distinguishability measure based on the error probability of discrimination when the same local measurement is performed on every copy.
We study some general properties of these measures,
including  the probability distribution of density matrices, defined via the volume element induced by the metric,  and illustrate their use in the paradigmatic cases of qubits and Gaussian infinite-dimensional states.

\end{abstract}

\maketitle

\section{Introduction}\label{intro}
About fifty years ago Herman Chernoff proved his famous bound,
which characterizes the asymptotic behavior of the minimal probability of error when discriminating  two hypothesis given a large number of observations~\cite{chernoff_measure_1952}.
Its quantum analog was recently conjectured~\cite{ogawa_error_2004} and finally proven
by combining the results of  two recent publications~\cite{audenaert_discriminating_2007,nussbaum_lower_2006}. In this
quantum setting one is confronted with the problem of knowing the
minimum error probability in identifying one of two possible known states of which $N$ identical copies are given.
Hereafter we will refer to this minimum  simply as the error probability~$P_{\rm e}$.
This problem is widely known as {\em quantum state discrimination}
\footnote{See~\cite{bergou_2004} and~\cite{chefles_2000} for two
reviews on the recent and more historical developments of this field
respectively.}. Its difficulty (but also its appeal) lies in the
fact that quantum mechanics only allows for full discrimination of
such states when they are orthogonal. This has both fundamental and
practical implications that lie at the heart of quantum mechanics
and its~applications.

For these past fifty years the classical Chernoff bound ---as well
as hypothesis testing in general--- has proved to be extremely
useful in all branches of science. Likewise, one would expect its
quantum version to be far more than a mere academic issue. The
characterization and control of quantum devices is a necessary
requirement for quantum computation and communication,  and quantum
hypothesis testing is specially designed for assessing the
performance of these tasks. Particularly important examples for
which state discrimination plays an essential role are quantum
cryptography~\cite{gisin_quantum_2002}, classical capacity of
quantum channels \cite{hayashi_general_2003}, or even quantum
algorithms~\cite{bacon_from_2005}.  Equally important are some new
theorems concerning different quantum extensions of hypothesis
testing: the quantum Stein's lemma, proved some years
ago~\cite{hiai_proper_1991,ogawa_strong_2000}, and the quantum
Hoeffding bound, recently established
in~\cite{nagaoka_converse_2006,hayashi_error_2006,audenaert_asymptotic_2007}.

In this paper we study the classical and the quantum Chernoff bounds
in connection to measures of distinguishability for quantum states,
putting special emphasis on the qubit and Gaussian cases. We start
by reviewing classical and quantum hypothesis testing and the
corresponding Chernoff  bounds in Sec.~\ref{CHT-CB} and
Sec.~\ref{QHT-QCB}, respectively (the latter includes the before
mentioned recent results by Nussbaum and Szkola \cite{nussbaum_lower_2006} and Audenaert {\em
et~al.} \cite{audenaert_discriminating_2007}). In Sec~\ref{DM} we discuss the notion of
a distinguishability measure for quantum states. We briefly motivate
an important instance of such a notion based on classical  statistical
measures, that is, the quantum fidelity, and move to a fully
operational alternative, based on the asymptotic rate exponent of
the error probability in symmetric quantum hypothesis testing: the
quantum Chernoff measure%
\footnote{By `operational' it is meant `defined though a specific procedure or task', in contradistinction to `purely mathematical'.  }.
We also discuss a similar
distinguishability measure derived from the same rate exponent  when
the decision is based on $N$ identical single-copy (local)
measurements ---instead of the collective measurements on the $N$
copies  assumed in the derivation of the quantum Chernoff bound. In
Sec.~\ref{M} we study the metrics induced by the previously defined
measures of distinguishability and give explicit expressions for general
$d$-dimensional systems. We also give the probability distribution
of the eigenvalues of a $d\times d$ density matrix based on the
quantum Chernoff metric (induced by the corresponding
distinguishability measure).  We find that the metric based on local
measurements is discontinuous and has to be defined piecewise: on
the set of pure states, where it agrees with the Fubini-Study
metric, and, separately, on the set of strictly mixed states, where
it agrees with one-half the Bures-Uhlmann metric.
 The quantum
Chernoff metric, in contrast, is  continuous 
and smoothly interpolates between the  Fubini-Study and
one-half the~Bures-Uhlmann metrics.
 In Sec.~\ref{QS}  we
concentrate on the particular case of two-level systems and study in
some depth the differences between the quantum Chernoff measure and
metric and those based on identical local measurements. In
Sec.~\ref{GS} we give explicit expressions of the quantum Chernoff
measure and its corresponding induced metric for general Gaussian
states. Finally, we state our conclusions  in~Sec.~\ref{C}.

\section{Classical hypothesis testing: Chernoff bound}\label{CHT-CB}

One of the most fundamental problems in statistical decision theory
is that of choosing between two possible explanations or models,
that we will refer to as hypothesis $H_{0}$ and $H_{1}$, where the
decision is based on a set of data collected from measurements or
observations. For example, a medical team has to decide whether a
patient is healthy (hypothesis $H_{0}$) or has certain disease
(hypothesis $H_{1}$) in view of  the results of some clinical test.
Often,~$H_{0}$ is called the working hypothesis or null hypothesis,
while $H_{1}$ is called the alternative hypothesis. In general these
two hypotheses do not have to be treated on equal footing, since
wrongly accepting or rejecting one of  them might have very
different consequences. These two types of errors, i.e., the
rejection of a true null hypothesis or the acceptance of a false
null hypothesis, are called type I or type II errors respectively,
and their corresponding probabilities will be denoted by
$p(1|H_0)\equiv p_0(1)$ and $p(0|H_1)\equiv p_1(0)$ throughout the
paper. In our example, failure to diagnose the disease is a type~II
error, whereas it is a type~I error to wrongly conclude that the
healthy patient has the disease. Of course it would be desirable to
minimize the two types of errors at the same time. However, this is
typically not possible since reducing  those of one type entails
increasing  those of the other type. Hence, a common way to proceed
is to minimize the errors of one type, while keeping those of the
other type bounded by a constant (which may depend on the number of
observations). Another (Bayesian-like) approach consists in
minimizing a linear combination of the two error probabilities
$P_{\rm e}=\pi_0 p(1|H_0) +\pi_1 p(0|H_1)$, where $\pi_0$ and
$\pi_1$ can be interpreted as the \emph{a priori} probabilities that
we assign to the occurrence of each hypothesis. In this paper we
consider this latter approach, which is known as
 \emph{symmetric} hypothesis testing.

For the sake of simplicity, we assume to start with
that~$\pi_0=\pi_1=1/2$, and we deal with tests that have only two
possible outcomes, $b=0,1$. This is, for example, the situation that
corresponds to the identification of a biased coin~that can be (with
equal probability) of one of two types: $0$ or $1$ (corresponding to
hypothesis $H_{0}$ or $H_{1}$ respectively). If it is of the type
$0$ the probabilities of obtaining head and tail are
respectively~$p_{0}(0)=p$ and~$p_{0}(1)=1-p\equiv \bar{p}$, while if
it is of type $1$ we write~$p_1(0)=q $ and
$p_1(1)=1-q\equiv\bar{q}$. The test consists in tossing the coin,
which has two possible outcomes: either head ($b=0$) or tail
($b=1$).

If we can toss the coin only once (single observation), it is easy
to convince oneself that the minimum (average) probability of error
is attained when we accept the hypothesis (decide that the tossed
coin is of the type) for which the observed outcome occurs with
largest probability. Therefore~\footnote{In this formula, as well as
in most of the formulas involving minimization throughout the paper,
one should properly write~$\inf_{s\in[0,1]}$ instead
of~$\min_{s\in[0,1]}$ since the minimum may not exist if~$p_0$
and~$p_1$ ($\rho_0$ and $\rho_1$ in the quantum case) are degenerate
and have different support. This is so because in this case the
continuity of the argument of~$\min_{s\in[0,1]}$ in all these
equations is  guaranteed {\em only} in the open interval $(0,1)$ and
(end-point) singularities may occur at~$s=0,1$. We will overlook
this mathematical subtlety in the main text  to simplify the
exposition.}
\begin{eqnarray}
\label{chernoff-classic-1}
   P_{\rm e}&=&\frac{1}{2}\sum_{b=0}^1 \min\{p_0(b),p_1(b)\}\nonumber\\
    &\leq& \frac{1}{2}\mathop{\min}_{s\in[0,1]} \sum_{b=0}^1
    p_0^s(b)p_1^{1-s}(b)\equiv P_{\rm CC},
\end{eqnarray}
where we have used the inequality $\min\{p,q\}\leq p^{s}q^{1-s}$.
The subscript CC stands for classical Chernoff. This expression also
holds for tests with more than two outcomes. We just need to extend
the sum over $b$ to the entire range of possible outcomes. In what
follows, we leave the range of $b$ unspecified whenever an
expression is valid for an arbitrary number of outcomes.

Next, let us assume we can toss the coin $N$ times. The set of
possible outcomes (the sample space) is the $N$-fold Cartesian
product of  $\{0,1\}$ (or~$\{{\rm head},{\rm tail}\}$). The two
probability distributions of these outcomes,~$p_0^{(N)}(b^{(N)})$
and~$p_1^{(N)}(b^{(N)})$, will be given by the product of the
corresponding single-observation
distributions,~$p^{(N)}_i(b^{(N)})=p_i(b_1)p_i(b_2)\cdots p_i(b_N)$,
where now~$b^{(N)}\equiv (b_1,b_2,\ldots ,b_N)\in\{0,1\}^{\times
N}$, and one immediately obtains~\cite{Cover:1991fj}
\begin{equation}\label{eq:chernoff-clasico}
    P_{\rm e}\leq \frac{1}{2}\mathop{\min}_{s\in[0,1]}\left( \sum_{b}
    p_0^s(b) p_1^{1-s}(b)\right)^N .
\end{equation}

This is the Chernoff bound \cite{chernoff_measure_1952}. It is
specially important because it can be proved to give the exact
asymptotic rate exponent of the error probability, that is,
\begin{eqnarray}\label{eq:chernoff-C}
    \displaystyle P_{\rm e}&\sim& {\rm e}^{-N C(p_0,p_1)};\nonumber \\
    \displaystyle C(p_0,p_1)&\equiv&-\mathop{\min}_{s\in[0,1]}\log \sum_{b}
    p_0^s(b) p_1^{1-s}(b) .
\end{eqnarray}

The so-called {\em Chernoff information}, or Chernoff
distance,~$C(p_0,p_1)$, can also be written in terms of the
 Kullback--Leibler divergence $K(p_{0}/p_{1})=\sum_{b}p_{0}(b)\log[{p_{0}(b)}/{p_{1}(b)}]$~\cite{Cover:1991fj}:
\begin{equation}\label{chernoff-binary}
    C(p_0,p_1)=K(p_{s^*}/p_{0})=K(p_{s^*}/p_{1}) ,
\end{equation}
where  
\begin{equation}
p_{s}(b)= {p_0^s(b) p_1^{1-s}(b)\over \displaystyle\sum_{b}
    p_0^s(b) p_1^{1-s}(b)};\quad s\in[0,1]
\end{equation}
is a family of probability distributions known as the Hellinger arc that
interpolates between $p_{0}$ and $p_1$, and $s^*$ is the value of
$s$ at which the second equality in~(\ref{chernoff-binary}) holds.
In other words, it is the point at which $p_s$ is equidistant to
both  $p_{0}$  and $p_{1}$ (in terms of Kullback--Leibler distance).
It can be shown that $s^*$ is also the value of $s$ that minimizes
the right hand side of~\eqref{eq:chernoff-C}.

\begin{figure}[htbp] %  figure placement: here, top, bottom, or page
   \centering
   \includegraphics[width=6 cm]{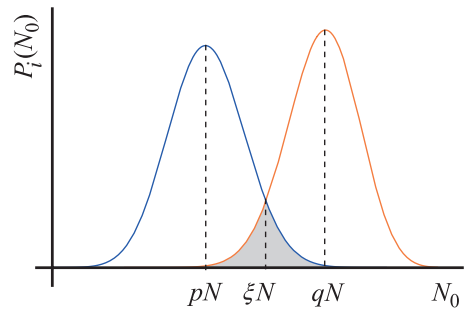}
   \caption{ (Color 
  online)   Each curve represents the probability to obtain~$N_0$ heads after $N$
   tosses of a bias coin that can be of one of two types, $0$ or $1$.
   The probability that the coin of type~$0$~($1$) produces a head at any given toss is $p$ ($q$).
For large~$N$ these curves approach Gaussian distributions centered
at~$pN$~and~$qN$, respectively. The point~$\xi N$ where they cross
defines the decision boundary (see main text). The error probability
is given by the shaded area.} \label{fig:gaussianitas}
\end{figure}

For the case of measurements with two outcomes, such as the example
of the coins discussed above, one can give a closed expression for
the Chernoff distance, which we denote in this binary case
as~$C(p,q)$:
 \begin{equation}
    C(p,q)=\xi \log\frac{\xi}{p}+\bar{\xi}\log\frac{\bar{\xi}}{\bar{p}},
\end{equation}
with
\begin{equation}\label{eq:chernoff-binary}
\xi\equiv {\log(\bar{q}/\bar{p})\over\log
(p/\bar{p})+\log(\bar{q}/q)};  \quad  \bar\xi\equiv 1-\xi.
\end{equation}
The parameter  $\xi$ has a very straightforward interpretation. If
$N_{0}$ is the number of heads (of $0$'s) after $N$ trials, which
according to the distribution~$p_0$ occurs with probability
\begin{equation}
P_0(N_0)={N \choose N_0} p^{N_0}{\bar p}\,{}^{N-N_0}
\end{equation}
[according to the distribution~$p_1$ it occurs with probability
$P_1(N_0)$, defined the same way but  with $p$ replaced by $q$],
then $\xi$ is the fraction of heads above which one must decide in
favor of~$p_0$. That is, if  $N_{0}\ge\xi N$ one accepts
hypothesis~$H_{0}$, while if $N_{0}<\xi N$ one accepts $H_{1}$.
Asymptotically, the contribution to the error probability is
dominated by situations where $N_{0}=\xi N$, i.e., by events that
occur with the same probability for both hypotheses (see
Fig.~\ref{fig:gaussianitas}). The probability of such events is
clearly a lower-bound to the probability of error. It is
straightforward to check that $-\lim_{N\to\infty}\log P_0(\xi N)/N$
[or equivalently $-\lim_{N\to\infty}\log P_1(\xi N)/N$] coincides
with the upper bound given by the Chernoff distance $C(p,q)$. This
proves that the Chernoff bound is indeed attainable.

\section{Quantum hypothesis testing: The~quantum Chernoff bound}\label{QHT-QCB}
We now tackle discrimination (symmetric hypothesis testing) in a
quantum scenario. We consider two sources,~$0$ and~$1$ that produce
states described respectively by the density matrices~$\rho_{0}$
and~$\rho_{1}$ acting on a Hilbert space~${\mathscr H}$. We are
given~$N$ copies of a state~$\rho$ with the promise that they have
been produced either by the source~$0$ (with prior
probability~$\pi_{0}$) or by the source~$1$ (with prior
probability~$\pi_{1}=1-\pi_{0}$). Accordingly, we can formulate two
hypothesis ($H_0$ and   $H_1$) about the identity ($0$ or $1$,
respectively) of the source that has produced these copies. We wish
to find a protocol to determine, with minimal error probability,
which hypothesis better explains the nature of the $N$ copies. No
matter how complicated this protocol might be, it is clear that the
output must be classical: we have to settle for one of the two
hypotheses. Therefore the protocol develops in two stages. First, to
obtain information about the states we must necessarily make a
(quantum) measurement, which in contrast to the classical world is
an inherently random and destructive process. Second, one has to
provide a classical algorithm that processes the measurement
outcomes (classical data) and produces the best answer ($H_0$ or
$H_1$).  Quantum mechanics allows for a convenient description of
this two-step process by assigning to each answer, $H_0$ and $H_1$,
a single POVM (positive operator valued measure) element $E_0$ and
$E_1$ respectively ($E_{b}\geq 0$ acts on ${\mathscr H}^{\otimes
N}$;  $E_{0}+E_{1}=\id$). The probability that this POVM measurement
gives the answer $H_b$ conditioned to $\rho=\rho_i$ is
$p_{i}(b)=\tr(\rho_{i}^{\otimes N}E_{b})$.

The problem thus reduces to finding the set of operators
$\{E_{b}\}_{b=0}^1$ that minimize the mean probability of error,
For the simplest case of a single copy ($N=1$) and two
equiprobable  hypotheses ($\pi_{0}=\pi_{1}=1/2$) it is \cite{Helstrom:1976kx}
\begin{equation}\label{Pe}
P_{\rm e}=\half \left[p_{0}(1)+p_{1}(0)\right]=\half\left[\tr (
\rho_{0} E_{1})+\tr ( \rho_{1} E_{0})\right].
\end{equation}
Since $E_{0}=\id-E_{1}$, we can introduce the {\em Helstrom matrix}
$\Gamma\equiv\rho_{1}-\rho_{0}$, as is common in quantum state
discrimination, and write
\begin{equation}
P_{\rm e}=\half-\half\tr\left(E_{1}\Gamma\right) ,
\end{equation}
which  only needs to be optimized with respect to~$E_{1}$.  We note
that $\Gamma$\ has some negative eigenvalues, as $\tr\, \Gamma=0$.
This necessarily implies that the minimum error probability is
attained if $E_{1}$ is the projector on the subspace of positive
eigenvalues of $\Gamma$. We will denote this projector by
$\{\Gamma>0\}$ and define the positive part of $\Gamma$
as~$\Gamma_{+}=\{\Gamma>0\}\Gamma$. Taking into account that $\Gamma
$ is traceless, we obtain
\begin{equation}
\tr(E_{1} \Gamma)= \tr\,\Gamma_{+}=\half \tr\,|\Gamma|,
\end{equation}
where the matrix $|A|$ (absolute value of $A$) is defined to be
$|A|=\sqrt{A^{\dag}
 A}$. We arrive at the final result \cite{Helstrom:1976kx},
\begin{equation}
P_{\rm e}=\half\left(1-\half\tr|\rho_{1}-\rho_{0}|\right)\mbox{.}
\end{equation}

The problem of discriminating multiple copies (arbitrary $N$) is thus formally
solved by replacing $\rho_i$ by $\rho_i^{\otimes N}$ in the above
equations. Indeed, if we do not have any restrictions on the type of
measurements performed on the~$N$ copies,~$E_1=\{\rho_0^{\otimes
N}-\rho_1^{\otimes N}<0\}$, and the mean probability of error
is~just
\begin{equation}\label{ebc02.08.07-1}
P_{\rm e}=\half\left(1-\half\tr\left|\rho_{1}^{\otimes
N}-\rho_{0}^{\otimes N}\right|\right)\mbox{.}
\end{equation}
However, the computation of the trace norm of the Helmstrom matrix
in~(\ref{ebc02.08.07-1}) is tedious and, moreover, this equation
provides little information about the large $N$ behavior of the
error probability, which is what the Chernoff bound is about.

The quantum version of the Chernoff (upper) bound was presented very
recently in~\cite{audenaert_discriminating_2007}. There it is shown
that
\begin{equation}\label{chernoff-quantum}
    P_{\rm e} \leq  
    {1\over2}\mathop{\min}_{s\in[0,1]} \tr \rho_0^s
   \rho_1^{1-s}  \equiv \pcher\equiv \half Q
\end{equation}
(the subscript~QC stands for quantum Chernoff), which holds for {\em
arbitrary} density matrices. Moreover, this bound can be very
efficiently computed.

The bound~(\ref{chernoff-quantum}) is a straightforward application
of the following theorem \cite{audenaert_discriminating_2007}:
\begin{theorem}\label{th:1}
Let $A$ and $B$ be two positive operators, then for all  $0\le s\le
1$,
\begin{equation}\label{eq:main}
\tr\left(A^s B^{1-s}\right) \ge
{1\over2}\tr\left(A+B-\left|A-B\right|\right).
\end{equation}
\end{theorem}
The proof of this theorem involves advanced methods in matrix
algebra and we refer the interested reader
to~\cite{audenaert_discriminating_2007}. Instead, here we will give
a simple proof of the inequality~\eqref{chernoff-quantum} where
instead of minimizing over~$s$, the particular value~$s=1/2$ will be
chosen.

We first notice that one obtains an upper-bound to~$P_{\rm e}$ by
picking any particular positive operator~$E_1$ (and, accordingly,
$E_0$)  in~\eqref{Pe}. A convenient choice is~$\tilde
E_{1}=\{\rho_{0}^{1/2}-\rho_{1}^{1/2}<0\}$  (and thus~$\tilde
E_{0}=\{\rho_{0}^{1/2}-\rho_{1}^{1/2}\geq0\}$), where, as above,
$\{A>0\}$ stands for the projector onto the subspace spanned by the
eigenstates of~$A$ with positive eigenvalue. After the following
series of inequalities we arrive to the desired
result~\cite{hayashi}:
\begin{eqnarray}
2 P_{\rm e}&\leq &\tr(\tilde E_{1}\rho_{0})+\tr(\tilde
E_{0}\rho_{1})
\label{ebc02.08.07-2} \\
&=& \tr(\rho_{0}^{1/2}\rho_{0}^{1/2}\{\rho_{0}^{1/2}-\rho_{1}^{1/2}<0\})+\nonumber\\
& &+\tr(\rho_{1}^{1/2}\rho_{1}^{1/2}\{\rho_{0}^{1/2}-\rho_{1}^{1/2}\geq0\}) \nonumber\\
&\leq& \tr(\rho_{0}^{1/2}\rho_{1}^{1/2}\{\rho_{0}^{1/2}-\rho_{1}^{1/2}<0\})+ \nonumber\\
& &+\tr(\rho_{0}^{1/2}\rho_{1}^{1/2}\{\rho_{0}^{1/2}-\rho_{1}^{1/2}\geq0\}) \nonumber\\
&=&
\tr[\rho_{0}^{1/2}\rho_{1}^{1/2}(\{\rho_{0}^{1/2}\!\!-\!\rho_{1}^{1/2}\!\!<0\}\!+\!
\{\rho_{0}^{1/2}\!\!-\!\rho_{1}^{1/2}\!\geq0\})] \nonumber\\
&=&  \tr(\rho_{0}^{1/2}\rho_{1}^{1/2}), \nonumber
\end{eqnarray}
where in the second inequality we have used
\begin{eqnarray}
(\rho_{1}^{1/2}-\rho_{0}^{1/2})\{\rho_{0}^{1/2}-\rho_{1}^{1/2}<0\}&\geq&
0;\nonumber \\
(\rho_{0}^{1/2}-\rho_{1}^{1/2})\{\rho_{0}^{1/2}-\rho_{1}^{1/2}\geq0\}&\geq&
0.
\end{eqnarray}
The general proof (for all~$s$) follows the same steps but
taking~$\tilde E_{1}=\{\rho_{0}^{1-s}-\rho_{1}^{1-s}<0\}$ if $0\le
s<1/2$ and~$\tilde E_{1}=\{\rho_{0}^{s}-\rho_{1}^{s}<0\}$ if $1/2<
s\le1$. In this case, the inequality analogous to the second one
in~(\ref{ebc02.08.07-2}) requires the two additional non-obvious
relations
\begin{eqnarray}
\kern-4em\tr\!\left[\rho_1^{1-s}\!\left(\rho_0^{s}-\!\rho_1^{s}\right)\!
\left\{\rho_0^{1-s}\!\!-\!\rho_1^{1-s}\!\ge\!0\right\}\right]\!\!
&\ge&\!\!0; \ 0\!\le \!s\!<\!\mbox{$\half$}\nonumber\\
\tr\!\left[\rho_0^s\!\left(\rho_1^{1-s}\!\!-\!\rho_0^{1-s}\right)
\!\left\{\rho_1^s-\!\rho_0^s\ge0\right\}\right]\!\!&\ge&\!\!0;
\ \mbox{$\half$}\!\le\! s\!<\!1.\phantom{....}
\end{eqnarray}
These inequalities follow immediately from the following non-trivial
lemma, which constitutes the core of the
proof~\cite{audenaert_discriminating_2007}:
\begin{lemma}\label{l:1}
Let $A$ and $B$ be two positive operators, then for all  $0\le t\le
1$, \be\label{eq:lemma} \tr[\{A-B\geq 0\} B (A^t- B^{t})] \ge 0. \ee
\end{lemma}

Before proceeding with the the asymptotic limit, several comments
about~\eqref{chernoff-quantum} are in order. (i)~The exponential
fall-off of the probability of error when a number $N$ of copies is
available follows immediately from $\tr(A\otimes B)=\tr A\;\tr B$:
\begin{equation}\label{ebc03.08.07-1}
P_{\rm e}\le {Q^N\over2}
\!={1\over2}\exp\left\{-N\!\left[-\!\min_{s\in[0,1]}\!\log\tr\rho_0^s\rho_1^{1-s}\right]\right\}.
\end{equation}
Remarkably enough, this rate exponent, which we may call {\em
quantum Chernoff information} because of its analogy with
$C(p_0,p_1)$, is asymptotically attainable, as follows from the
results of~\cite{nussbaum_lower_2006}. This is the quantum extension
of the classical result~(\ref{eq:chernoff-C}) and was first
conjectured by Ogawa and Hayashi in~\cite{ogawa_error_2004}. (ii)~If
the two matrices $\rho_{0}$ and $\rho_{1}$ commute the bound reduces
to the classical Chernoff bound \eqref{chernoff-classic-1}, where
the two probability distributions are given by the spectrum of the
two density matrices. (iii)~The function~$Q_{s}= \tr
\rho_{0}^s\rho_{1}^{1-s}$  (whose minimum gives the best bound) is a
convex function of~$s$ in~$[0,1]$, which means that a stationary
point will automatically be the global minimum
(see~\cite{audenaert_discriminating_2007} for a proof). This is a
very useful fact when computing the quantum Chernoff
bound~(\ref{chernoff-quantum}). (iv)~$Q$ is jointly concave in
$(\rho_{0},\rho_{1})$, unitarily invariant, and non-decreasing under
trace preserving quantum
operations~\cite{audenaert_discriminating_2007}. (v)~The quantum
Chernoff bound gives a tighter bound than that given by the {\em
quantum fidelity}
\begin{equation}\label{fidedef}
F(\rho_0,\rho_1)\equiv
\left(\tr\sqrt{\sqrt\rho_0\,\rho_1\sqrt\rho_0}
\,\right)^2\!\!=\left(\tr\left|\sqrt\rho_0\sqrt\rho_1\right|
\right)^{2},
\end{equation}
which is the most widely used quantum distinguishability measure
(see next section). This follows from the following set of
inequalities:
\begin{equation}\label{ebc07.08.07-1}
P_{\rm e}\!\leq\!\pcher\!\le\!{
\tr\rho_0^{1\!/2}\rho_1^{1\!/2\!}\over2}\!\leq \!
{\tr\!\left|\sqrt\rho_0\sqrt\rho_1\right|\over2}\!=\!
{\sqrt{F(\rho_0,\rho_1)}\over2}.
\end{equation}
In fact, the fidelity also provides a lower-bound to the probability
of error~\cite{fuchs_cryptographic_1999}:
\begin{equation}\label{ebc07.08.07-1'}
{1-\sqrt{1-F(\rho_0,\rho_1)}\over2}\le P_{\rm e} .
\end{equation}
In the case where one of the states (say $\rho_{0}$) is pure the
upper bound to the error probability can be made
tighter~\cite{nielsen_quantum_2000,kargin_chernoff_2005}:
\begin{equation}\label{ebc07.08.07-3}
P_{\rm e}\le \pcher= {Q\over2}  ={1\over2}F(\rho_0,\rho_1) .
\end{equation}
(vi)~The quantum Chernoff bound can be easily extended to the case
where the two states $\rho_0$ and $\rho_1$ (sources) are not
equiprobable:
\begin{equation}\label{chernoff-quantumgeneral}
    P_{\rm e} \leq  %\pcher  \equiv
    \mathop{\min}_{s\in[0,1]}\pi_{0}^s \pi_{1}^{1-s} \tr \rho_{0}^s
   \rho_{1}^{1-s} .
\end{equation}
(vii)~The permutation invariance of the $N$-copy density matrices,
$\rho_i^{\otimes N}$, guarantees that the optimal collective
measurement can be implemented efficiently (with a polynomial-size
circuit known as {\em quantum Schur
transform})~\cite{bacon_efficient_2006}, and hence that the minimum
probability of error is achievable with reasonable resources.

As stated above, for multiple-copy discrimination the error
probability decreases exponentially with the number~$N$ of copies:
$P_{\rm e}\sim \exp\left[{-N D(\rho_0,\rho_1)}\right]$ as $N$ goes
to infinity~\cite{Cover:1991fj}. The error (rate) exponent
$D(\rho_0,\rho_1)$ is defined generically by
\begin{equation}\label{distance}
    D(\rho_0,\rho_1)=-\lim_{N\to\infty}\frac{1}{N}\log P_{\rm e}
\end{equation}
and characterizes the asymptotic behavior of the error probability.
From~(\ref{ebc03.08.07-1}) we readily see that if the best (joint)
measurement is used  it coincides with the quantum Chernoff
information,
    \begin{equation}\label{lambda-chernoff}
     D_{\rm QC}(\rho_0,\rho_1)=-\min_{s\in[0,1]} \log  \tr\rho_0^s
   \rho_1^{1-s} ,
 \end{equation}
where the equality holds because of the attainability
of~(\ref{ebc03.08.07-1}) discussed above and we have added
the subscript~QC. Moreover, this asymptotic value is also attained
by the square root (or ``pretty-good'') measurement
(see~\cite{kholevo_asymptotically_1979,hausladen_classical_1996} for
the precise definition). This immediately follows from the known
bounds~\cite{barnum_reversing_2002,harrow_many_2006}~$P_{\rm e}\leq
P_{\rm e}^{\rm SRM} \leq 2 P_{\rm e}$, where~$P_{\rm e}^{\rm SRM}$
is the error probability of discrimination when the square root
measurement is used.

Before closing this section, we briefly come back to the fidelity
bounds in~(\ref{ebc07.08.07-1}--\ref{ebc07.08.07-3}) and simply note
that the first two inequalities translate into the following bounds
to the rate exponent:
\begin{equation}\label{ebc09.08.07-1}
 -{1\over2} \log F(\rho_0,\rho_1)\le D_{\rm QC}(\rho_0,\rho_1)\le -\log F(\rho_0,\rho_1).
\end{equation}
If one of the states is pure Eq.~(\ref{ebc07.08.07-3}) implies that
the factor~$1/2$ in~(\ref{ebc09.08.07-1}) becomes~$1$ and we have
the exact relation
\begin{equation}\label{ebc09.08.07-1'}
D_{\rm QC}(\rho_0,\rho_1)=-\log F(\rho_0,\rho_1).
\end{equation}

\section{Distinguishability measures}\label{DM}

In this section we aim to define a measure of distinguishability
between states using the results reviewed in
Sec.~\ref{QHT-QCB}. Before doing so we will briefly outline how
classical statistical methods can be used to (partially) accomplish
this goal. We will then discuss an operational measure of
distinguishability based on the error probability in multiple-copy
state discrimination, leading to the quantum Chernoff measure.
Finally we will define the analogous quantity for local
discrimination protocols.

\subsection{Classical statistical approach}

The notion of distance between states is a fundamental issue that
has been studied for a long time. A straightforward way to
define such a distance is to take any suitable norm in the space of
states. However, a more physical approach, kick-started by the
pioneering work in~\cite{wootters_statistical_1981}, is to relate
the inherently probabilistic nature of quantum measurements to
classical statistical measures of distinguishability between
probability distributions.

 In particular, the author in~\cite{wootters_statistical_1981} uses the notion of \emph{statistical distance},
 \begin{equation}
 d_{\rm S}(p_0,p_1)=\arccos \sqrt{{\mathscr F}(p_0,p_1)},
 \end{equation}
as a measure of distinguishability between the probability
distributions $p_0$ and $p_1$, where
\begin{equation}
{\mathscr F}(p_0,p_1)=\left(\sum_{b} \sqrt{ p_0(b) p_1(b)}\right)^2
\end{equation}
is the \emph{statistical fidelity}. Accordingly, he defines a
distinguishability measure between quantum states~$\rho_{0}$
and~$\rho_{1}$ by maximizing~$d_{\rm S}(p_0,p_1)$
[i.e.,~minimizing~${\mathscr F}(p_0,p_1)$] over all possible POVM
measurements, characterized  by all possible sets of operators
$\{E_{b}\}_{b=1}^M$ with outcome probabilities given by
$p_0(b)=\tr(E_{b}\rho_{0})$ and $p_1(b)=\tr(E_{b}\rho_{1})$.
The statistical distance as such makes sense only when the number of
samplings of the probability distribution is large. Hence, in the
quantum extension of this notion it is implicitly assumed that one
performs the  {\em same} measurement on each of a large number~$N$
of copies of the state~$\rho\in\{\rho_0,\rho_1\}$. The optimization
over such  local repeated  measurements leads to one of the most
widely used distinguishability measures \cite{fuchs96distinguishability}:  The (quantum)
fidelity~$F(\rho_0,\rho_1)$, defined in~(\ref{fidedef}).

The fidelity, or statistical distance, has many desirable
properties:  (i)~it is easily computable;  (ii)~for pure states it
reduces to the standard distance given by the angle between rays in
the Hilbert space~$\mathscr H$;  (iii)~as mentioned above, it
provides bounds to~$P_{\rm e}$. Nevertheless, a strict physical interpretation
is so far unclear, and its definition is based on repeated local
measurements, while quantum mechanics allows for much more general
ways to access the information contained in the $N$ copies, via
collective measurements on the whole of them.

\subsection{Quantum Chernoff distance}

A very natural and also operational distinguishability measure is
provided by the error probability of discrimination. As a first
candidate, one could take this very error probability $P_{\rm e}$
for a given fixed number~$N$ of copies. However, the  choice of a
particular~$N$ in such a definition would not only be arbitrary but
also problematic since one can find examples~\cite{Cover:1991fj}
where~$P_{\rm e}(\rho_{0},\rho_{1};N)>P_{\rm
e}(\rho_{0}',\rho_{1}';N)$, whereas $P_{\rm
e}(\rho_{0},\rho_{1};M)<P_{\rm e}(\rho_{0}',\rho_{1}';M)$  for a
different number~$M$ of copies. A straightforward way to go around
this problem is to use the asymptotic expressions for $N\to\infty$
and define the distinguishability measure as the largest rate
exponent in~(\ref{distance}). We further note that the presence of the logarithm 
ensures that~$D(\rho_0,\rho_1)=0$ if and only if~$\rho_0=\rho_1$,
while the minus sign makes distinguishability decrease as
discrimination becomes more difficult, i.e., as $P_{\rm e}$
increases.

The quantum Chernoff information,~$D_{\rm QC}(\rho_0,\rho_1)$, is
therefore a physically meaningful  and efficiently computable
distinguishability measure. Note that~\eqref{lambda-chernoff}  does
not \textit{stricto sensu} define a distance, since it does not
fulfil the triangular inequality. It has however all of the other
properties that one should expect from a reasonable measure.
This, in itself, is already a remarkable fact since, as far as
measures and metrics are concerned, there is usually a compromise
among operational definiteness, computability and
contractivity~\cite{gilchrist_distance_2005}. For instance, the
distance proposed in~\cite{lee_operationally_2003}, although having
an operational definition, is not contractive.

We point out that another operational distinguishability measure can be obtained in asymmetric hypothesis testing by minimizing the type II error rate while keeping  the type I error rate upper-bounded by a fixed value.
The optimal  error rate in this situation is provided by the quantum Stein's Lemma  \cite{hiai_proper_1991, ogawa_strong_2000} and leads to the well known quantum relative entropy.
Despite of having an operational meaning, the quantum relative entropy has two obvious drawbacks as a distinguishability measure: it is not symmetric on its arguments and it diverges if one of the states is pure.
 
\subsection{Classical Chernoff distance: local measurements}\label{sub-CCD:LM}

In the derivation of the quantum Chernoff bound one optimizes  over
all possible quantum measurements, in particular over quantum joint
measurements on ${\mathscr H}^{\otimes N}$, that act over all
the~$N$ copies coherently.
It is of great interest, both theoretically and in practice, to know
whether such  joint measurements are strictly necessary to attain
the bound or one can make do with separable ones (which include
those that can be implemented with local operations and classical
communication, simply known as LOCC measurements). As far as we are
aware, the answer to this is unknown. This question is also relevant
in connection with the operational meaning attached
to~$D(\rho_0,\rho_1)$. In this section we focus on this operational aspect and
compute~$D(\rho_0,\rho_1)$ from its definition in~(\ref{distance})
assuming that the discrimination protocol $P_{\rm e}$ refers to is
constrained to make use of the same individual measurements, defined
by a local POVM $\{E(b)\}_{b=1}^M$, on each of the~$N$ available
copies. We loosely refer to these protocols as  \emph{local}. 
Local protocols are relevant from the theoretical point of view since they help to elucidate the 
role of quantum correlated measurements in asymptotic hypothesis testing. 
For example, in quantum phase estimation
local measurements suffice to achieve the collective bounds~\cite{kholevo_asymptotically_1979}.  Here, we will show that these protocols do not achieve the quantum Chernoff bound. In addition, from a more practical  point of view, local protocols are much simpler to implement experimentally, specially in a situation where the number of sub-systems is increasingly  large.

In such a local protocol, after the measurements have been performed
we have a sample of~$N$ elements of the probability distribution
$p_{i}(b)=\tr(E_{b}\rho_{i})$, $i=0,1$,  based on which we have to
discriminate between the candidate~$H_{0}$ or~$H_{1}$. In such
a scenario the error probability, which we call~$P^{\rm loc}_{\rm e}$,
can be obtained using the classical Chernoff
bound~\eqref{chernoff-classic-1} applied to the
distributions~$p_{0}$ and~$p_{1}$.  One can thus define the error
exponent~(\ref{distance}) and thereby introduce a new operational
distinguishability measure based on local discrimination:
\begin{equation}\label{eq:Dcc}
D_{\rm CC}(\rho_0,\rho_1)=-\min_{\{E_{b}\}}\min_{s\in[0,1]} \log
\sum_{b}p_{0}^s(b) p_{1}^{1-s}(b),
 \end{equation}
where the subscript ${\rm CC}$ reminds us that we have made use of
the classical Chernoff bound.

The measure~$D_{\rm CC}(\rho_0,\rho_1)$ is obtained by maximizing
the rate exponent over all possible single-copy generalized
measurements~$\{E_{b}\}_{b=1}^M$ (just as is done for the fidelity).
Unfortunately, there is no simple closed expression for this maximum
for general mixed states. However, we do encounter again the
relation~(\ref{ebc07.08.07-1}) with the fidelity: since the square
root of the statistical fidelity ${\mathscr F}(p_0,p_1)$ upper
bounds~$P_{\rm CC}$ in~(\ref{chernoff-classic-1}), it also upper
bounds the local error probability~$P^{\rm loc}_{\rm e}$. That is,
\begin{equation}\label{ebc07.08.07-2'}
P^{\rm loc}_{\rm e}\!\leq\! P_{\rm CC}\!\leq \!\min_{\{E_{b}\}}\!{
\sqrt{{\mathscr F}(p_0,\!p_1\!)}\over2}\!= \!
{\sqrt{F(\rho_{0},\!\rho_{1}\!)}\over2},
\end{equation}
and
\begin{equation}\label{ebc07.08.07-2}
D_{\rm CC}(\rho_0,\rho_1) \ge- {1\over2}\log F(\rho_0,\rho_1).
\end{equation}

Since $D_{\rm QC}(\rho_0,\rho_1)\ge D_{\rm CC}(\rho_0,\rho_1)$, we
note that whenever  $D_{\rm QC}(\rho_{0},\rho_{1})=-(1/2) \log
F(\rho_{0},\rho_{1})$ the inequality~(\ref{ebc07.08.07-2}) has to be
saturated. This, in turn, means that in this situation one can
optimally discriminate between~$H_0$ and~$H_1$ just by performing
a fixed local measurement on each of the~$N$ copies (no collective
measurements are required to attain the quantum Chernoff bound).

There is still another important situation when the quantum Chernoff
bound is attainable by local measurements: when one of the states (say
$\rho_0$) is pure. If this is the case, Eq.~(\ref{ebc07.08.07-3})
holds and $D_{\rm QC}(\rho_0,\rho_1)=-\log F(\rho_0,\rho_1)$. To
prove that $D_{\rm CC}(\rho_0,\rho_1)=D_{\rm QC}(\rho_0,\rho_1)$,
let us consider the two-outcome measurement defined by $E_0=\rho_0$,
$E_1=\openone-\rho_0$. Note that $p_0(1)=\tr (E_1\rho_0)=0$
and \hbox{$p_0(0)=\tr (E_0\rho_0)=1$}. After performing this measurement on
each of the $N$ copies the protocol proceeds as follows: we accept
$H_0$ if all of the outcomes are~$0$, otherwise we accept~$H_1$. One
may refer to this classical data processing as {\em unanimity vote} \cite{acin_multiple-copy_2005}.
The error probability can be easily computed by noticing that no
error occurs unless we get~$N$ times the outcome~$0$
[since~$p_0(1)=0$]. Therefore,
\begin{equation}\label{ebc07.08.07-4}
P^{\rm loc}_{\rm e}\!=\!\pi_1 p^N_1\!(0)\!=\!\pi_1\left[
\tr\left(\rho_0\rho_1\right)\right]^N\!=\!\pi_1
\left[F(\rho_0,\rho_1)\right]^N ,
\end{equation}
where the last equality holds because $\rho_0$ is assumed to be a
pure state. From this equation it follows immediately that $D_{\rm
CC}(\rho_0,\rho_1)\!=\!-\log F(\rho_0,\rho_1)=\!D_{\rm
QC}(\rho_0,\rho_1)$, and the quantum Chernoff bound is attainable by
local measurements.
It also follows from the first equality in~(\ref{ebc07.08.07-4})
that this result corresponds to taking the limit $s\!\to\!0$
in~(\ref{chernoff-classic-1}).

\section{Metric}\label{M}

The set of states of a quantum system, as  that of classical probability distributions on a given sample space,\footnote{For sake of clarity, in this section we assume a finite sample space, but the results hold also for general probability measures over continuous spaces.}  can be endowed with a metric structure~\cite{bengtsson_geometry_2006}, and thus thought of as a Riemannian manifold. This enables us to relate geometrical concepts (e.g., distance, volume, curvature, parallel transport) to physical ones (e.g.,  state discrimination and estimation, geometrical phases). Among the novel applications of metrics in quantum information, they have been recently used to characterize quantum phase transitions~\cite{zanardi_bures_2007}.

The first step towards this geometric approach to quantum states is to define  the line element $ds$ or (infinitesimal) distance between two neighboring ``points" $\rho$ and $\rho-d\rho$. All local properties follow from this definition. More precisely, they follow from the metric, i.e., from the set of coefficients of $ds^2$ when written as a quadratic form in the differentials of the coordinates (parameters) that specify the quantum states. There is, however, no unique choice of $ds$ unless some monotonicity conditions are invoked. 

For classical probability distributions, $\{p(b)\}$, a line element is singularized  (up to a propotionality factor) by imposing that it be
non-increasing under stochastic maps. It is the well known Fisher metric (in what follows the terms metric and line element will be used interchangeably):
\begin{equation}
ds_{\rm F}^2= \frac{1}{4}\sum_{b} \frac{[dp(b)]^2}{p(b)}%
.
\label{fisher}
\end{equation}

In contrast to the classical case, the monotonicity condition under completely positive (quantum stochastic) maps does not define a metric uniquely, which explains why a substantial body of research
on quantum metrics has emerged over the last years. Among the main developments, Petz~\cite{petz_monotone_1996} has characterized the family of quantum contractive metrics by establishing a correspondence with operator-monotone functions.

An alternative, more physical approach is to  
define a line element from a suitable distinguishability measure between infinitesimally close states. A remarkable example is given in~\cite{braunstein_statistical_1994}. In  this seminal paper Braunstein~and Caves consider a one-parameter family of states $\rho(\theta)$ and~map the problem of distinguishability to that of estimating the parameter~$\theta$  optimally. They define a line element, $ds_{\rm BC}^2$, as $d\theta^2$ expressed in the appropriate units of statistical deviation (roughly speaking, $d\theta^2$ divided by the minimal error in the estimation of $\theta$). 
By making use of classical statistical methods (Cram\'er-Rao bound) they
find
\begin{equation}
\label{eq:BC}
ds^2_{\mathrm{BC}}= 4 \max_{\{E_{b}\}} ds^2_{\rm F}= \max_{\{E_{b}\}} I_{\rm F} \;d\theta^2,
\end{equation}
where $I_{\rm F}=\sum_b [dp(b)/d\theta]^2/p(b)$ (it is the so called Fisher information), with $p(b)=\tr [E_b \rho(\theta)]$, and the maximization is over all possible POVM measurements $\{E_{b}\}$ on a single copy of $\rho(\theta)$. They also succeed in giving a closed expression for $ds^2_{\mathrm{BC}}$ and
show that their metric coincides up to a factor with that induced by the Bures-Uhlmann distance~\cite{Bures_1969,uhlmann_transition_1976}
\begin{equation}
\label{buresdistance}
d_{\rm
BU}(\rho_0,\rho_1)=\sqrt2\left[1-\sqrt{F(\rho_0,\rho_1)}\right]^{1/2}.
\end{equation}
More precisely, they show that  $ds^2_{\mathrm{BC}}=4 ds^2_{\rm BU}$, where
\begin{equation}
ds^2_{\rm BU}  \equiv  [d_{\rm BU}(\rho,\rho-d\rho)]^2 
\label{ebc5.12.07-1}
\end{equation} 
[see also~\eqref{eq:dsbu} below] and a series expansion to $O(d\rho^2)$ is understood in the right hand side of this equation. 
We note in passing that for commuting states, i.e.,  classical probability distributions, the Bures-Uhlmann line element $ds^2_{\rm BU}$ coincides with the Fisher metric \eqref{fisher}. A~quantum metric with such normalization is said to be Fisher adjusted. 

Although one can obtain a finite distance $d_{\rm BC}(\rho_0,\rho_1)$ for arbitrary states $\rho_0$ and $\rho_1$ by integrating $ds_{\rm BC}$ along geodesics, it is important to notice that the operational meaning of the Braunstein and Caves metric is lost in the process.

In the spirit of Braustein and Caves' physical approach to metrics, we next consider the distinguishability measures $D_{\rm QC}$ and $D_{\rm CC}$, discussed in Section~\ref{DM}, for infinitesimally close states and derive line elements with the same operational meaning, which we call $ds_{\rm QC}$ and $ds_{\rm CC}$ respectively. For $ds_{\rm QC}$ we also give the volume element and the prior probability distribution, whereas those corresponding to the metric $ds_{\rm CC}$ can be easily found in the literature since, as will be shown, $ds^2_{\rm CC}$ is proportional to the widely-studied Bures metric $ds^2_{\rm BU}$.

Before we start 
we would like to point out that one could also consider line elements induced by other quantities, such as the quantum relative entropy, which, as we saw above, also has a clear operational interpretation.  The quantum relative entropy induces  the so-called  Kubo-Mori metric~\cite{petz_covariance_2002}, which has the drawback of being singular for pure states.

\subsection{Quantum Chernoff metric}

For neighboring density matrices $\rho$ and $\rho-d\rho$ (e.g., those for which their
independent matrix elements differ by an infinitesimal amount) the distinguishability
measure~$D(\rho,\rho-d\rho)$ defines a metric, as in  \eqref{ebc5.12.07-1}.
For the quantum Chernoff measure, $D_{\rm QC}$, this metric can be computed from
Eq.~(\ref{lambda-chernoff})~\cite{audenaert_quantum_2006}:

\begin{equation}
ds^2_{\rm QC}=1-\min_{s\in[0,1]}\tr[\rho^s(\rho-d\rho)^{1-s}]+\dots
, \label{ds2a}
\end{equation}
where the dots stand for higher order terms in $d\rho$ that will not
contribute to $ds^2$ and we have also used that~$\log y=y-1+\dots$. We
now recall the integral representation
\begin{equation}
a^t = \frac{\sin(t\pi)}{\pi} \int_0^{\infty} dx\,\,
\frac{ax^{t-1}}{a+x};\quad 0<t<1 \label{e1}
\end{equation}
and its derivative,
\begin{equation}
\label{e2} t a^{t-1} = \frac{\sin(t\pi)}{\pi} \int_0^{\infty}
dx\,\, \frac{x^{t}}{(a+x)^2}; \quad -1<t<1 .
\end{equation}
These representations hold for $a> 0$ and can be
straightforwardly extended to positive matrices.
In particular, using \eqref{e1} and the convergent sequence
\begin{equation}
{1\over a-b}=a^{-1} +a^{-1} b a^{-1}+a^{-1} b a^{-1} b
a^{-1}+\ldots,
\end{equation}
which also holds for matrices provided $a> b$, one can write, up to
second order in $d\rho$,
\begin{eqnarray}
(\rho\!-\!d\rho)^{1-s}\!\!\!\!&=&\!\!\! c_{s} \int_0^{\infty}
\!\!dx\,(\rho\!-\!d\rho) \frac{x^{-s}}{\rho-d\rho+x}
\\
\!\!\!&\approx&\!\!\! c_{s} \int_0^{\infty} \!\!dx\,x^{-s} (\rho\!-\!d\rho)\left(\frac{1}{\rho+x}\right.\nonumber\\
\!\!\!&+&\!\!\!\left.\frac{1}{\rho+x}d\rho\frac{1}{\rho+x}+
\frac{1}{\rho+x}d\rho\frac{1}{\rho+x}d\rho\frac{1}{\rho+x}\!\right),\nonumber
\end{eqnarray}
where $c_{s} =\pi^{-1}\sin(s\pi)$.
Inserting this expansion in \eqref{ds2a} one finds
\begin{eqnarray}
ds^2_{\rm QC}&=&\max_{s\in(0,1)} c_{s} \int_0^{\infty} dx \tr\left[
\frac{x^{1-s}}{(\rho+x)^2}\rho^sd\rho\right.
\nonumber \\
&&\left.+
\frac{x^{1-s}}{(\rho+x)^2}\rho^sd\rho\frac{1}{\rho+x}d\rho\right].
\end{eqnarray}
The first term in the integrand vanishes, as can be seen by using
\eqref{e2} and $\tr\, d\rho=0$, while the second term can be
computed in the eigenbasis~$\{\ket{i}\}$ of~$\rho$;
$\rho=\sum_{i}\lambda_{i}\ketbrad{i}$:
\begin{eqnarray}
ds^2_{\rm QC}\!\!\!&=&\!\!\!\!\!\max_{s\in(0,1)}  \sum_{ij} c_{s}
\int_0^{\infty} dx x^{1-s} \frac{\lambda_{i}^s\, |\langle
i|d\rho|j\rangle
|^2}{(\lambda_{i}+x)^2(\lambda_{j}+x)}\nonumber\\
\!\!\!&=&\!\!\!{1\over2}\max_{s\in(0,1)}  \sum_{ij} \frac{|\langle
i|d\rho|j\rangle |^2}
{(\lambda_{i}-\lambda_{j})^2}\!\left(\lambda_{i}\!+\!\lambda_{j}\!-\!\lambda_{i}^{s}
\lambda_{j}^{1-s}\!\!-\!\lambda_{j}^{s}\lambda_{i}^{1-s}\right)\nonumber\\
\!\!\!&=&\!\!\!\frac{1}{2} \sum_{ij} \frac{|\langle i|d\rho|j\rangle
|^2}{(\lambda_{i}-\lambda_{j})^2}\!\left(\lambda_{i}+\lambda_{j}-2\sqrt{\lambda_{i}\lambda_{j}}\right)
,
\end{eqnarray}
where in the second equality we have taken into account that
$d\rho=d\rho^\dagger$, which enabled us to  symmetrize the
expression in parenthesis  that multiplies $|\langle
i|d\rho|j\rangle |^2$ in the sum (this symmetrization gives the
factor $1/2$). 
The quantum Chernoff metric can be finally written as,
\begin{equation}
ds^2_{\rm{QC}}=
\frac{1}{2}\sum_{ij} \frac{|\langle i|d\rho|j\rangle
|^2}{(\sqrt{\lambda_{i}}+\sqrt{\lambda_{j}})^2}. \label{eq:dsqc}
\end{equation}

The quantum Chernoff metric  belongs to
the family of contractive quantum metrics, as it should, since by construction the probability of error  cannot be improved by a pre-processing of the states. In
fact the quantum Chernoff metric coincides with a member of this
family that has been explicitly written by Petz in~\cite{D.Petz:fk}
and with the so called Wigner-Yanase metric, which has been recently studied in depth by the authors of~\cite{gibilisco_wigner-yanase_2003}. In particular, 
the geodesic distance,
the geodesic path, and the scalar curvature of the quantum Chernoff metric can be read off from their Eqs.~(5.1-5.3).

 By separating diagonal from off-diagonal terms, the metric in~(\ref{eq:dsqc}) can also be written as
 \begin{equation}
ds^2_{\rm QC}= \sum_{i} \frac{(d\lambda_i)^2}{8\lambda_{i}}
+\sum_{i<j} \frac{|\langle
i|d\rho|j\rangle|^2}{(\sqrt{\lambda_{i}}+\sqrt{\lambda_{j}})^2}.
\label{ebc12.07.07}
\end{equation}
Next, we wish to identify the degrees of freedom in the off-diagonal
terms. We will see that they correspond to infinitesimal unitary
transformations acting on $\rho$ (which leave its eigenvalues
unchanged).  
This is most conveniently done by parameterizing~$\rho$ by its eigenvalues  and eigenvectors, namely by~$\lambda_i$ and the components of $\ket i$ onto a given canonical basis $\{\ket{\alpha_k}\}$:
\begin{equation}
U_{k i}\equiv\langle \alpha_k |i\rangle=\bra{\alpha_k}U\ket{\alpha_i}
\label{ebc12.12.07-1}
\end{equation}
(naturally, it also holds that $U_{k i}=\bra{k}U\ket{i}$).
A neighboring density matrix $\rho'=\sum_i\lambda'_i\ket{i'}\bra{i'}$  is thus
parameterized by  $\lambda'_i=\lambda_i+d\lambda_i$ and    $U'_{k i}=U_{k i}+dU_{k i}=\langle\alpha_k|i'\rangle$. We further note that   
$
\ket{i'}=(\openone+\delta T)
\ket i 
$,
where $\delta T$ is antihermitian, $\delta T^\dagger=-\delta T$.
It is actually  the
infinitesimal generator along the direction in parameter space that
takes $\{\ket{i}\}$ into~$\{\ket{i'}\}$. It follows that $dU_{k i}=\bra{\alpha_k}\delta T\ket i $. The matrix elements of
$d\rho$ can be expressed~as
\begin{eqnarray}
\bra i d\rho\ket j&\!\!\!=\!\!&\bra i(\rho'-\rho)\ket
j=\sum_{k}\langle i|k'\rangle\langle k'|j\rangle
\lambda'_k-\lambda_i \delta_{ij}
\nonumber\\
&\!\!\!=\!\!& d\lambda_i\delta_{ij} +(\lambda_j-\lambda_i)\bra
i\delta T\ket j + O(\delta T^2) ,
\end{eqnarray}
and those of $\delta T$ as
\begin{eqnarray}
\bra i\delta T\ket j&=&\sum_k \langle i|\alpha_k\rangle \bra{\alpha_k}
\delta T\ket j=\sum_k U^*_{ki} dU_{kj}\nonumber\\
&=&\sum_k\bra{\alpha_i}U^\dagger\ket{\alpha_k} \bra{\alpha_k}dU\ket{\alpha_j}\nonumber\\
&=&\bra{\alpha_i}U^\dagger dU\ket{\alpha_j}\equiv
 \left(U^\dagger dU\right)_{ij},
\label{ebc13.07.07-2}
\end{eqnarray}
where we have used~(\ref{ebc12.12.07-1}) in going from the first to the second line [the very same matrix elements of $\delta T$ can also be written as $(dU\, U^\dagger)_{ij}$ in the eigenbasis of $\rho$].
Substituting these relations back into~(\ref{ebc12.07.07}) we obtain
\begin{equation}
ds^2_{\rm
QC}\!=\!\!\sum_i\!{(d\lambda_i)^2\over8\lambda_i}+\!\sum_{i<j}
\left(\!\sqrt{\lambda_i}-\!\sqrt{\lambda_j}\right)^2\!\left|\left(U^{\dagger}dU\right)_{ij}
\right|^2 . \label{ebc13.07.07}
\end{equation}
The same expression can also be derived by differentiating 
\begin{equation}
\rho=U^\dagger\rho^{(0)}\, U ,
\end{equation}
where $\rho^{(0)}\equiv \sum_i \lambda_i \ket{\alpha_i}\bra{\alpha_i}$
is diagonal in the canonical basis and has the spectrum of $\rho$.

Eq.~(\ref{ebc13.07.07}) displays  the metric  $ds^2_{\rm QC}$ in a very
suggestive form. Any density matrix can be parameterized  by its
eigenvalues $\{\lambda_i\}$ and the unitary matrix $U$ that
diagonalizes it. Eq.~(\ref{ebc13.07.07}) expresses the infinitesimal
distance between two such matrices in terms of these very parameters. The
first term is immediately recognized as the (Fisher) metric on the
$(d-1)$-dimensional simplex of eigenvalues of $\rho$, which is
assumed to be $d\times d$ throughout the rest of this section (note
that $\sum_i\lambda_i=1$, which implies~$\sum_i d\lambda_i=0$).
Thus, {\em stricto senso}, it should be expressed in terms of a set
of~$d-1$ independent eigenvalues. If we choose this~set to
be~$\{\lambda_i\}_{i=1}^{d-1}$ the first term in~(\ref{ebc13.07.07})
becomes
\begin{equation}
{1\over8} \sum_{i,j}^{d-1} g_{\rm F}^{ij}d\lambda_i d\lambda_j  ,
\label{ebc13.07.07-4}
\end{equation}
where the subscript~$\rm F$ stands for Fisher, and
\begin{equation}
g_{\rm
F}^{ij}\!=\!{\delta^{ij}\over\lambda_i}+{\Phi^{ij}\over1\!-\!\sum_i^{d-1}\lambda_i};\
\mbox{$\Phi^{ij}\!=1$ for $\!1\le i,j\le d\!-\!1$}.
\end{equation}
It follows that the determinant of $g_{\rm F}$, which we will need
below, is
\begin{equation}
\det g_{\rm F}
={(\lambda_1\cdots\lambda_{d-1}\lambda_d)^{-1}}  .
\label{ebc13.07.07-3}
\end{equation}

The second term in~(\ref{ebc13.07.07}) contains the
factors~$|(U^{\dagger}dU)_{ij} |^2$, which are invariant under
left-multiplication [since the left-hand side
of~(\ref{ebc13.07.07-2}) is independent of the choice of basis
$\{\ket\alpha\}$]. Hence, the {\em normalized} volume element
induced by these terms will coincide with the (unique) Haar
measure~$dV_{\rm H}$ of~$U(d)/[U(1)]^d$, known as the {\em flag
manifold} $Fl_{\mathbb C}^{(d)}$ (see
e.g.,~\cite{zyczkowski_induced_2001} and references therein). Using
the wedge product of differential forms, this Haar measure can be
written as
\begin{equation}\label{ebc05.08.07-1}
dV_{\rm H}=\frac{1}{C_{\rm H}}\left|\bigwedge_{i<j}\mathrm{Re}
(U^{\dagger}dU)_{ij} \wedge \mathrm{Im}(U^{\dagger}dU)_{ij} \right|,
\end{equation}
where $C_{\rm H}$ is a normalization constant so that \hbox{$\int
dV_{\rm H}=1$.} Note that the one-form basis
in~(\ref{ebc05.08.07-1}) contains $2\times [d(d-1)/2]$ (real and
independent) elements, which indeed coincides with  the $d^2-d$
independent parameters of~$U(d)/[U(1)]^d$.

Volume elements (derived from metrics) are of great interest because
they give a canonical way of defining prior probability
distributions on continuous sets. According to this approach,
Eqs.~(\ref{ebc13.07.07}--\ref{ebc05.08.07-1}) provide a means to
define such probability distribution for general density matrices: if
$\thb=(\theta_1,\theta_2,\dots)$ is a set of independent real parameters that specifies the density matrices
as $\rho(\thb)$ and the metric is written as $ds^2=d\thb{\gb}\,d\thb^t$ (i.e., $\gb$ is the metric tensor), then we can define the prior ${\mathcal P}[\rho({\thb})] $ through the relation
${\mathcal P}[\rho({\thb})] \,\prod_\al d\theta_\alpha=dV/\int dV$,
where $dV= \sqrt{\det {\gb}}\,\prod_\al d\theta_\alpha$. It follows
from~(\ref{ebc13.07.07}) that ${\mathcal P}[\rho({\thb})]$ is the
product of two independent probability distributions: one that
depends  exclusively on the parameters encoded in the unitary
matrix~$U$ and expresses the fact that
they  are simply distributed according to the Haar measure~$dV_{\rm
H}$; and one, denoted as~${\mathscr P}(\{\lambda_i\})$, that gives
the probability distribution of eigenvalues. The latter can be
written as
\begin{equation}\label{ebc06.08.07-1}
{\mathscr P}(\!\{\lambda_i\}\!)\!=\!{1\over C_d}\!\prod_{i}^d
\!{1\over\sqrt{\lambda_i}}\,\delta\!\!\left(\!\mbox{$1\!-\!\sum_j\lambda_j$}\!\right)
\!\prod_{i<j}\!\left(\!\!\sqrt{\lambda_i}-\!\sqrt{\lambda_j}\right)^2\!\!
,
\end{equation}
where for a given dimension $d$ the constant $C_d$ is chosen to
ensure that probability adds up to one.

The prior distribution on the simplex  of eigenvalues of~$\rho$ for
the Bures metric (see below), analogous to~${\mathscr
P}(\{\lambda_i\})$ in~(\ref{ebc06.08.07-1}), was proposed
in~\cite{Hall:1998qy}, but it took considerable efforts to compute
the right normalization constant. Slater \cite{slater-1999-32} gave
values for dimensions~\hbox{$d= 3 , 4 , 5$} and finally Sommers
and \.Zyczkowski \cite{sommers-2003-36} managed to give a general
expression for arbitrary finite dimensions. Here we will compute
$C_d$ following similar techniques.

The coefficient~$C_d$ is defined by the normalization
condition~$\int {\mathscr P}(\{\lambda_i\})\prod_i^d d\lambda_i=1$.
Thus,~$C_d=I(1)$, where
\begin{equation}
I(r)\!=\!\!\int_0^\infty \!\prod_{i}^d\!
{d\lambda_i\over\sqrt{\lambda_i}}\,\delta\!\!\left(\!\mbox{$r^2\!\!-\!\sum_j\lambda_j$}\!\right)
\!\prod_{i<j}\!\left(\!\sqrt{\lambda_i}-\!\sqrt{\lambda_j}\right)^2\!\!
.
\end{equation}
Although we only need this integral for $r=1$, the introduction of
this radial parameter~$r$ enables us to compute the normalization
$I(1)$ more easily. We first note that by re-scaling $\lambda_i\to
r^2\lambda_i$ one gets
\begin{equation}
I(r)=r^{d^2-2} I(1)
\end{equation}
[i.e., $I(r)$ is a homogeneous function of $r$ of degree $d^2-2$],
and thus
\begin{equation}
\int_0^\infty dr\, r \,{\rm e}^{-r^2} I(r)= I(1) \int_0^\infty dr\,
r^{d^2-1} {\rm e}^{-r^2} .
\end{equation}
It follows from this equation that
\begin{eqnarray}
C_d=I(1)&=&{2^{d}\over\Gamma(d^2/2)}\int_0^\infty \prod_{i}^d
{d\lambda_i\over2\sqrt{\lambda_i}}
\;{\rm e}^{-\sum_i\lambda_i}\nonumber\\
&\times&
\prod_{i<j}\left(\!\sqrt{\lambda_i}-\!\sqrt{\lambda_j}\right)^2 .
\end{eqnarray}
This expression can be further simplified by the change of variables
$\lambda_i\to t_i=\sqrt{\lambda_i}$, which leads to
\begin{equation} \label{cd}
C_d={2^{d}\over\Gamma(d^2/2)}\int_0^\infty \prod_{i}^d {dt_i} {\rm
e}^{-t_i^2} \,\prod_{i<j}\left(t_i-t_j\right)^2 .
\end{equation}
By expanding the square of the Vandermonde determinant $\prod_{i<j}
(t_i-t_j)$, one could in principle compute~$C_d$ in terms of Euler
gamma functions. However this is very impractical  since the number
of terms in such an expansion grows exponentially with~$d$. A much more
efficient way to proceed is as follows. Let $\{P_k(t)=a_k
t^k+a_{k-1} t^{k-1}+\dots+a_1 t+a_0\}$,
$a_k\not=0$, be a family or orthonormal polynomials in the
set~$[0,\infty)$ with a weight function of Hermite type, so that
\begin{equation}
\label{ebc15.08.07-1} \int_0^\infty dt\,{\rm
e}^{-t^2}P_k(t)P_l(t)=\delta_{kl} .
\end{equation}
Note that~$\{P_k(t)\}$ are not Hermite polynomials, since the
integration range is $[0,\infty)$ instead of $(-\infty,\infty)$. Now, if we define the renormalized polynomials $Q_k(t)\equiv P_{k}(t) /a_k$  it
is not hard to show that
\begin{equation}
\prod_{i<j}(t_i-t_j)\!=\!\left|
\matrix{Q_{d-1}(t_1)&Q_{d-2}(t_1)&\dots&Q_0(t_1)\cr
Q_{d-1}(t_2)&Q_{d-2}(t_2)&\dots&Q_0(t_2)\cr
\vdots&\vdots&\ddots&\vdots\cr
Q_{d-1}(t_d)&Q_{d-2}(t_d)&\dots&Q_0(t_d)} \right|  .
\end{equation}
Substituting in to ~(\ref{cd}) and using
the orthonormality of $P_k$, one has
\begin{equation} \label{cd'}
C_d={2^{d} d!\over\Gamma(d^2/2)} \prod_{k=0}^{d-1} a_k^{-2}  .
\end{equation}
In contrast to the examples considered in
Ref.~\cite{sommers-2003-36}, and as far as we are aware, there is no
known closed expression for the leading coefficients~$a_k$ for the
case at hand. However, Eq.~(\ref{cd'}) provides an efficient way of
computing the quantum Chernoff normalization constant~$C_{d}$; e.g.,
by applying the Gram-Schmidt  orthogonalization algorithm [with the
internal product defined in Eq.~(\ref{ebc15.08.07-1})] one easily
obtains the coefficients~$a_k$, and thereby~$C_d$ . We give the
value of this constant for~$d\le6$:
\begin{eqnarray}
C_2&=& \pi-2;\nonumber\\
C_3&=& {8\over35}(\pi-3);\nonumber\\
C_4&=&{6\pi^2-29\pi+32\over 6720};\nonumber\\
C_5&=&{128(72\pi^2-435\pi+656)\over21082276215};\nonumber\\
C_6&=&{9(480\pi^3-3747\pi^2+9352\pi)-65536\over2023466257612800} .
\end{eqnarray}

\subsection{Classical Chernoff/Bures metric}\label{sub-CC:BM}

{}From the local measure $D_{\rm CC}(\rho_0,\rho_1)$,
Eq.~(\ref{eq:Dcc}), one  can readily obtain the corresponding
\emph{local} metric. If $0\!<\!p~(b)\!=\tr( \rho E_{b})<1$  for every
measurement outcome~$b$, direct differentiation of $D_{\rm
CC}(\rho,\rho-d\rho)$ leads to
\begin{equation}
\label{eq:dscc}
ds^2_{\rm CC}={1\over2} \max_{\{E_b\}} \;ds^2_{\rm F} ,
\end{equation}
where $ds^2_{\rm F}$ is the Fisher metric~(\ref{fisher}),
with~$p({b})=\tr(\rho E_{b})$, $dp(b)=\tr(d\rho E_{b})$ and
$s^*=1/2$ being the value of~$s$ that achieves this minimum
in~(\ref{eq:Dcc}). The maximization of \eqref{fisher} over the local measurements
$\{E_b\}_{b=1}^M$, which commutes with the minimization over~$s$ as
long as~$p(b)\neq 0,1$,  results
in~\cite{braunstein_statistical_1994}
\begin{equation}
ds_{\rm{BU}}^2=
 \frac{1}{2}
\sum_{ij} \frac{|\bra id\rho\ket j|^2}{\lambda_{i}+\lambda_{j}},
\label{eq:dsbu}
\end{equation}
or equivalently,
\begin{equation}
ds^2_{\rm
BU}=\sum_i{(d\lambda_i)^2\over4\lambda_i}+\sum_{i<j}{(\lambda_i-\lambda_j)^2\over
\lambda_i+\lambda_j}\left|(U^\dagger dU)_{ij}\right|^2,
\label{ebc08.08.07-1}
\end{equation}
where we use the same notation as in~(\ref{eq:dsqc})
and~(\ref{ebc13.07.07}), respectively. This is  the  Bures-Uhlmann
metric, which, as mentioned above, can be also obtained from the Bures
distance~\eqref{buresdistance}~\cite{hubner_explicit_1992}.  
{}From~\eqref{eq:dscc}
we then have
\begin{equation}\label{bures-metric}
ds_{\rm{CC}}^2 = \half  ds_{\rm BU}^2= \half [1-F(\rho,\rho-d\rho)]
\end{equation}
for strictly mixed states (the last equality holds to
order~$d\rho^2$). The
corresponding prior probability distribution (quantum Jeffreys
prior)  was derived and calculated
in~\cite{Hall:1998qy,slater-1999-32,sommers-2003-36}.

If one of the states is pure (say $\rho_{0}$, as in previous
sections) then the classical distribution $p(b)$ becomes degenerate
[$p(0)=1$] for the optimal choice~$E_0=\rho_{0}$ (recall the last
comments in Sec.~\ref{sub-CCD:LM}), and the previous derivation does
not hold. In this case, the optimal choice of~$s$
in~(\ref{chernoff-classic-1}) is obtained by taking the limit
$s\to0$, as we already discussed in~Sec.~\ref{sub-CCD:LM}. Recalling
the first equality in~(\ref{ebc07.08.07-4}), we obtain
$D_{CC}(\rho,\rho-d\rho)=-\log[p(0)-dp(0)]=dp(0)$ [note that
$dp(0)\ge0$ since $1\ge p(0)-dp(0)=1-dp(0)$], which is linear in
$dp(b)$ and therefore does not define a proper metric in probability
space. From the results of~Sec.~\ref{sub-CCD:LM} we also know that
if one of the states is pure then   $D_{\rm CC}(\rho_0,\rho_1)=-\log
F(\rho_0,\rho_1)$ and therefore
\begin{equation}\label{bures-metric2}
ds_{\rm{CC}}^2 =1-F(\rho,\rho-d\rho)=    ds_{\rm BU}^2
\end{equation}
for pure states. This agrees with the previous discussion since
$dp(0)=1-F(\rho,\rho-d\rho)$ if $\rho$ is a pure state.
Eq.~(\ref{bures-metric2})~has to be taken with special care. It
gives a valid metric for the set of pure states (which {\em only}
includes variations in the unitary parameters), i.e., when
$\rho-d\rho$ is also a pure state ($\rho-d\rho=U\rho U^\dagger$).
Moreover, for pure states~$ds_{\rm{CC}}^2$ coincides with the
Fubini-Study metric [recall that the Bures-Uhlmann metric is
Fubini-Study adjusted~\cite{sommers-2003-36}, hence this statement
follows from Eq.~(\ref{bures-metric2})].

By combining Eqs.~(\ref{bures-metric}) and~(\ref{bures-metric2}), we
see that $ds_{\rm CC}^2$ shows a discontinuity when the mixed state
$\rho$ approaches the set  of pure states.  The quantum Chernoff
metric~\eqref{eq:dsqc} does not have this pathology.
This can be seen by comparing \hbox{the
$i<j$}  ($d\lambda_i=0$) terms in~(\ref{ebc13.07.07}) with those
in~(\ref{ebc08.08.07-1}) (the diagonal terms~$i=j$ coincide). As
$\lambda_j\to \delta_{1j}$ ($\rho$ approaches a pure state), we
readily see that $ds^2_{\rm QC}\to ds^2_{\rm BU}$. In the opposite
situation, when~$\rho$ approaches the completely mixed state
$\openone/d$, we can write~$\lambda_i=1/d +\epsilon_j$, where
$\epsilon_j$ approaches zero. Expanding the~$i<j$ terms in
both~(\ref{ebc13.07.07}) and~(\ref{ebc08.08.07-1}) we can check that
$ds^2_{\rm QC}=\half ds^2_{\rm BU}$ up to terms of order $\epsilon^3$.
We conclude that the quantum Chernoff metric smoothly interpolates
between the two components (that on strictly mixed states and that
on pure states) of the local metric~$ds^2_{\rm CC}$.  We will come
back to this point in the next section, where qubit states are
discussed as an example to illustrate the results in this and in
previous sections.

\section{Qubit states}\label{QS}
In this section we apply our results to qubit mixed states, that is, general
two-dimensional states. We will first study the distinguishability measures 
$D_{\rm QC}$ and $D_{\rm CC}$ and then move on to the corresponding metrics and priors.

For qubits one has $\rho_i=(\openone+ \vec{r}_i\cdot\vec{\sigma})/2$, $i=0,1$,
where $\vec r_i$ is the Bloch vector of $\rho_i$,  $0\le|\vec
r_i|\equiv r_i\le1$. The eigenvalues of $\rho_i$ are $\wp_i=(1+
r_{i})/2$ and $\bar \wp_i\equiv 1-\wp_i$.  It is straightforward to
obtain
\begin{eqnarray}\label{chernoff-qubits}
Q_{s} &\equiv& \tr \rho_0^s
  \rho_1^{1-s}=  \left( {\wp_{0}}^s {\wp_{1}}^{1-s}+{\bar \wp_{0}}{}^s {\bar \wp_{1}}{}^{1-s}\right)
  \cos^2{\theta\over2}\nonumber\\
  &&+ \left( {\wp_{0}}^s {\bar \wp_{1}}{}^{1-s}+{\bar \wp_{0}}{}^s {\wp_{1}}{}^{1-s}\right)
  \sin^2{\theta\over2}  ,
\end{eqnarray}
where~$\theta$ is the angle between~$\vec{r}_0$ and $\vec{r}_1$. The
value of~$s$ that minimizes $Q_{s}$ and hence
gives~(\ref{chernoff-quantum}) and~\eqref{lambda-chernoff} is in
general a function of~$r_i$ and~$\theta$.  However, one can check
that in the particular case $r_0=r=r_1$ the minimum is at~$s^*=1/2$
\footnote{Qubit states are an example for which
the doubly stochastic matrix $D_{ij}=|\bra{i}U\ket{j}|^2$
is symmetric ($D_{ij}=D_{ji}$). Therefore, for isospectral states,
$Q_{s}(\rho,U\rho\, U^\dagger)=\sum_{ij}\lambda_{i}^s
\lambda_{j}^{1-s} D_{i,j}=\sum_{ij}(\lambda_{i}^s
\lambda_{j}^{1-s}+\lambda_{j}^s \lambda_{i}^{1-s}) D_{ij}$, which
has its minimum at~$s^*=1/2$.}.

\begin{figure}[ht]
\setlength{\unitlength}{1cm}
\begin{picture}(8,4.6)
\put(3.2,0.2){\epsfxsize=200pt\epsffile[90 3 363 171]{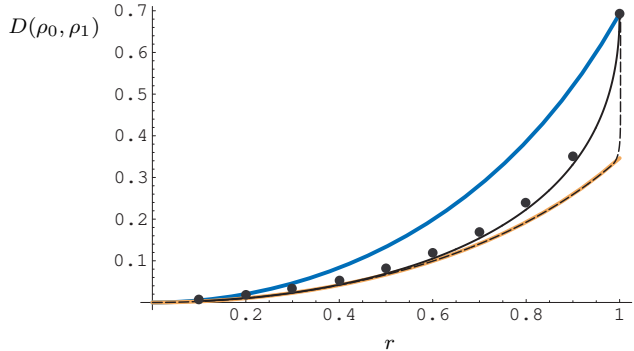}}
\put(-0.4,4){\footnotesize $D(\rho_{0},\rho_{1})$}
\put(4.6,-0.2){\footnotesize $r$}
\end{picture}
\caption[]{\label{fig1}(Color online) Measures of distinguishability between two-qubit states with relative angle $\theta=\pi/2$ for different values of $r=r_0=r_1$:  Values extrapolated from exact evaluation of the probability of error for $30\le N\le 35$ (dots);  bounds provided by the fidelity, 
Eq.~(\ref{ebc09.08.07-1}) (shaded); measure based on identical local
measurements, i.e., $D_{\rm CC}(\rho_0,\rho_1)$ (dashed); measure
based on collective measurements, i.e., $D_{\rm QC}(\rho_0,\rho_1)$
(solid line). }\label{fig:D}
\end{figure}

In Fig.~\ref{fig1} we plot the quantum Chernoff distinguishability
measure~$D_{\rm QC}(\rho_{0},\rho_{1})$ and the measure based on
local measurements $D_{\rm CC}(\rho_{0},\rho_{1})$ together with the
bounds~(\ref{ebc09.08.07-1}) provided by the fidelity, for states of
equal purity $r_0=r_1=r$ and for  $\theta=\pi/2$.
Notice that in general local measurements perform much worse than
the collective ones and~$D_{\rm CC}(\rho_0,\rho_1)$ runs remarkably
close to (actually, coincides with) the fidelity
lowerbound~(\ref{ebc09.08.07-1}) for most values of~$r$. However, as
it approaches the pure-state regime ($r\to 1$) it rapidly increases
towards its upper-bound. The reason for this rapid change can be
understood by recalling the unanimity vote protocol discussed
in~Sec.~\ref{sub-CCD:LM}. For two pure states,
$\rho_i=\ketbrad{\psi_i}$  (as corresponds to $r=1$), it boils down
to~\cite{acin_multiple-copy_2005} projecting along one of the
states, say~$\ket{\psi_0}$, and its
orthogonal,~$\ket{\psi_0^\perp}$. After performing this measurement
on each of the $N$ copies, if all of them project on $\ket{\psi_0}$,
one claims that the unknown state is $\ket{\psi_0}$ (hypothesis
$H_0$). However, if at least {\em one} of them projects on
$\ket{\psi_0^\perp}$ the guess is $\ket{\psi_1}$ (one accepts
$H_1$). This corresponds to $\xi=1$ in~(\ref{eq:chernoff-binary}).
For pure states it reaches the joint-measurement Chernoff bound by
making use of a much less demanding local-measurement protocol (see
also~\cite{brody_minimum_1996,acin_multiple-copy_2005} for the
optimal local strategy for finite~$N$).

In contrast, near the completely mixed state~$\openone/2$, for low
$r$, the optimal local strategy consists in choosing the measurement
$\{E_{0},E_{1}\}$ such that
$p=p_{0}(0)=\tr(\rho_0E_{0})=\tr(\rho_1E_{1})=p_1(1)=\bar q$, with
$p>1/2$. In this case, the acceptance of either $H_0$ or $H_1$ is
done on the basis of  a {\em majority vote} protocol: $H_0$ is
accepted if the outcome~$0$ occurs more times than the outcome~$1$ does,
i.e, $N_0=N/2$ [see also Eq.~(\ref{eq:chernoff-binary})]. It follows
from~(\ref{chernoff-binary}) \hbox{that $s^*=1/2$.} Therefore, the
lower-bound provided  by the fidelity, Eq.~(\ref{ebc09.08.07-1}), is
saturated [$s=s^*=1/2$ saturates the second inequality
in~(\ref{ebc07.08.07-2'}) and thus it also
saturates~(\ref{ebc07.08.07-2})]. This protocol  is optimal up to a
given value of the purity, i.e., for  $r\le r^*(\theta)$. For larger
values of~$r$ the `voting rule' (given by $\xi$) starts changing and
so does~$s^*$. Accordingly,~$D_{\rm CC}(\rho_0,\rho_1)$ moves away
from its lower-bound to end up saturating its upper bound at~$r=1$.

We next consider the metrics induced by local and  by joint
measures. The former, in particular, requires special attention
because of the abrupt behavior of $D_{\rm CC}(\rho_0,\rho_1)$ near
the set of pure states.  Indeed the critical value $r^*(\theta)$,
beyond which majority vote is no longer optimal,   goes to one as
the relative angle~$\theta$ between the Bloch vectors of the states
becomes smaller; $r^{*}(\theta)\rightarrow1$ as $\theta\to0$. As a
result, the sudden increase of~$D_{\rm CC}(\rho_1,\rho_2)$ develops
into a jump discontinuity at~$r=1$  [from~$-(1/2) \log
F(\rho_0,\rho_1)$ if~$r<1$ to~$-\log F(\rho_0,\rho_1)$ if~$r=1$].
For this reason, when defining the corresponding metric we have to
distinguish these two regions:   the set  of strictly mixed states
($r<1$) and the set of pure states ($r=1$).

In the region $r<1$ the outcome probabilities will never be
degenerate and the metric reduces to the Fisher metric, which upon
optimization over local measurements coincides with one-half the
Bures metric:
\begin{equation}\label{CCmetric}
ds_{\rm CC}^2={1\over2}ds^2_{\rm BU}=\frac{1}{8}\left(
\frac{dr^2}{1-r^2}+r^2 d\Omega^2, \right),
\end{equation}
 where $d\Omega^2=d\theta^2+\sin^2\theta d\phi^2$ is the usual metric on the 2-sphere.

 In the region~$r=1$
(pure states), the before-mentioned unanimity vote protocol is
optimal and the resulting metric is
\begin{equation}\label{CCmetric'}
ds^2_{\rm CC}={1\over4}d\Omega^2=ds^2_{\rm FS},
\end{equation}
where $ds^2_{\rm FS}$ is the well known  Fubini-Study metric, which,
as mentioned above,
also coincides with the Bures metric $ds^2_{\rm BU}$ in the limiting
case~$r\to1$. We notice again that $ds^2_{\rm CC}$ in
Eq.~(\ref{CCmetric'}) is a factor $2$ larger than
$\lim_{r\to1}ds^2_{\rm CC}$ in Eq.~(\ref{CCmetric}), where the limit
is taken along the lines~${dr=0}$.  The local distinguishability
measure thus induces a discontinuous metric or, phrased in a
different way, two different metrics for pure states or for strictly
mixed states.

This can be visualized using the Uhlmann representation, that is,
by embedding the Bloch sphere~$r\le 1$ in~${\mathbb R}^4$. To this
end,  one simply needs to define the new coordinate as $t=\cos\tau$,
where $\sin \tau\equiv r$. In spherical~coordinates one has
\begin{equation}
ds_{\rm CC}^2=\left\{
\begin{array}{lcc}
\displaystyle{1\over8}\left(d\tau^2+\sin^2 \tau\, d\Omega^2\right);&\ &0\le\tau < \pi/2\\[1em]
\displaystyle{1\over4} d\Omega^2;&&\tau=\pi/2
\end{array}
\right.
\end{equation}
where the first line correspond to strictly mixed states and the
second to pure states. We note that in the second (first)  line
$ds_{\rm CC}^2$ is nothing but the standard metric on a 2-sphere
(the top half of a 3-sphere)  of radius  $2^{-1}$ ($2^{-3/2}$).

\begin{figure}[ht]
\setlength{\unitlength}{1cm}
\begin{picture}(8,5)
\put(1.3,1.3){$A$} \put(0.2,1.3){$B$} \put(1.3,3.6){$C$}
\put(0,0){\epsfxsize=180pt\epsffile[ 0 3 168 119]{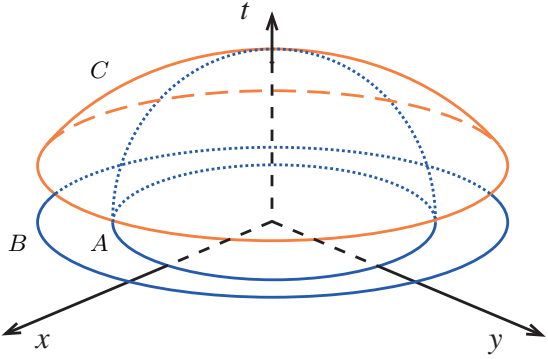}}
\end{picture}
\caption[]{\label{fig2} (Color 
online) Uhlmann representation of the set of single
qubit states according to metric $ds^2_{\rm CC}$,  based on $P^{\rm loc}_{\rm e}$
for local repeated measurements ($A$ and $B$), and according to the
quantum Chernoff metric $ds^2_{\rm QC}$, based on $P_{\rm e}$ for  general joint
measurements ($C$).}
\end{figure}

In Fig. \ref{fig2}, $A$  and $B$ represent (the slice $z=0$ of)
these two manifolds. One readily  sees that  the radius  of $B$
(pure states) is a factor $\sqrt2$ larger than that of the limiting
circle of~$A$ (for $r\to1\ \Leftrightarrow \ t\to0$, i.e.,
$\tau\to\pi/2$).

The quantum Chernoff (collective-measurement based)  metric can be
readily obtained from~\eqref{lambda-chernoff}
[or~\eqref{ebc13.07.07} particularized to qubit mixed states]:
\begin{equation}\label{metric=chernof}
   ds^2_{\rm QC}=\frac{1}{8}\left[ \frac{dr^2}{1-r^2}+2\left(1-\sqrt{1-r^2}\right)d\Omega^2
   \right].
\end{equation}
This metric quantifies distinguishability of qubit states in a
precise and operational way, and encapsulates the full power of
quantum mechanics.  It approaches the  Fubini-Study metric $ds^2_{\rm
FS}$ for pure states and also $ds^2_{\rm CC}$
for very mixed states, i.e. for small $r$. The metric smoothly
interpolates between the two regimes. By defining $r\equiv \sin 2
\tau$ with $0\le\tau\leq \pi/4$ we obtain again the standard metric
on a 3-sphere but this time of radius $1/\sqrt2$:
\begin{equation}
ds_{\rm QC}^2=\frac{1}{2}\left(d\tau^2+\sin^2 \tau d\Omega^2\right).
\end{equation}
The corresponding manifold is denoted by $C$ in Fig.~\ref{fig2}.
Geometrically the space of states endowed with the quantum Chernoff
metric $ds^2_{\rm QC}$ is a spherical cap defined by~$0\le\tau\leq
\pi/4$ whose radius is twice that of the Bures-like hemisphere~$A$.
In order to emphasize that the two metrics, are equal up to
order~$r^3$ at~$\tau\approx 0$, i.e.,~$r\approx 0$
(near~$\openone/2$), in the figure we have shifted the center of the
larger sphere so as to make the two manifolds tangent at~$\tau=0$.
The fact that $ds^2_{\rm CC}=\half ds^2_{\rm BU}=ds^2_{\rm
QC}+O(r^4)$ is a particular example of a general relation that we
discussed at the end of Sec.~\ref{sub-CC:BM}.

{}From the quantum Chernoff metric one can obtain a proper finite
distance (satisfying the triangle inequality) by, for example, computing
the geodesic distance,
\begin{equation}
\kern-.5em  d_{\rm QC}(\rho_0,\!\rho_1\!)\!= \!\! {\arccos(\!
\cos\tau_0 \cos\tau_1 \cos\theta \!+ \!\sin\tau_0\sin\tau_1
)\over\sqrt{2}},
\end{equation}
where $r_{i}\equiv \sin2\tau_{i}$ and $\theta$ is the relative angle
between the respective Bloch vectors.

The volume element and the prior distribution of density matrices
for qubit mixed states, which we here denote as~${\mathcal
P}[\rho(\vec r)]$, can be easily obtained from the above metrics.
According to the local and quantum Chernoff~metrics we have
respectively:
\begin{eqnarray}
{\mathcal P}_{\rm CC}[\rho(\vec r)]&=&\frac{\sin\theta}{\pi^2}\frac{r^2}{\sqrt{1-r^2}},\\
{\mathcal P}_{\rm QC}[\rho(\vec
r)]&=&\frac{\sin\theta}{2\pi(\pi-2)}\frac{1-\sqrt{1-r^2}}{\sqrt{1-r^2}},
\end{eqnarray}
where it is understood that~$r$ and~$\theta$ are the length and the
azimuthal angle of the Bloch vector of~$\rho$. Since the Haar volume
density on the 2-sphere is $\sin\theta/(4\pi)$, we see that the
eigenvalues of $\rho$, $\lambda_{\pm}=(1\pm r)/2$ are distributed
according to
\begin{eqnarray}
{\mathscr P}_{\rm CC}(\lambda_{\pm})&=&\frac{4}{\pi}\frac{r^2}{\sqrt{1-r^2}},\\
{\mathscr P}_{\rm
QC}(\lambda_{\pm})&=&\frac{2}{\pi-2}\frac{1-\sqrt{1-r^2}}{\sqrt{1-r^2}}.
\end{eqnarray}
(One can check that the latter agrees with our results in Sec.~\ref{M}.)
This have been recently used in~\cite{de_burgh_choice_2007}  to
assess the accuracy of different quantum tomographic measurements.

\section{Gaussian States}\label{GS}

We now illustrate  our results with  infinite-dimensional systems.
In particular we will focus on the family of single-mode Gaussian
states. This is a very significant class of quantum states  mainly
for two reasons. First, it has a very simple mathematical
characterization that allows for the derivation of  otherwise highly
non-trivial results, and, second, it describes accurately states of
light that are realized with current technology. In the following we
show that the Quantum Chernoff information, besides being the
natural distinguishability measure,  has the advantage of being
relatively easy to compute. The calculation of the fidelity, for
instance, is much more involved, as is apparent
from~\cite{abe_estimation_1999,twamley_bures_1996,paraoanu_bures_1998,wang_bures_1998,slater_quantum_1996},
where one can find such calculations for different classes of
gaussian states.

Gaussian states are by definition those that have a gaussian
characteristic function. The (symmetrically ordered) characteristic
function of one such state, $\rho$, is:
\begin{equation}
\chi(u)\equiv \tr[{\mathscr D}(u) \rho]= \exp\!\left(\!-i u^t \sigma
\xi -\frac{1}{4} u^t \sigma^t\Gamma\sigma u\!\right),
\label{charact}
\end{equation}
where $t$ denotes transposition,  $\sigma $  is the symplectic
matrix
\begin{equation}
\sigma=\pmatrix{0&1\cr -1&0}
\end{equation}
and $\mathscr D(u)=\exp[i (u_{2} \hat{q}-u_{1}\hat{p})]$  is the
displacement operator,  with $u=(u_{1},u_{2})^t$ and with position
and momentum operators satisfying~$[\hat{q},\hat{p}]=i$. The
annihilation and creation operators, defined as
$a=(\hat{q}+i\hat{p})/\sqrt{2}$ and
$a^\dagger=(\hat{q}-i\hat{p})/\sqrt{2}$, fulfil the canonical
commutation relations. The positivity of $\rho$ implies that the
$2\times2$ covariance matrix $\Gamma$ is real-symmetric and
satisfies  $\Gamma+i\sigma \geq 0$. A symplectic transformation is a
linear transformation  $S^t(\hat{q},\hat{p})$ that preserves the
commutation relations, or more succinctly  $S\sigma S^t=\sigma$.
Under such a transformation the displacement vector $\xi=(q, p)^t$
and the covariance matrix transform as  $\tilde{\xi}=S\xi$ as
$\tilde{\Gamma}= S\, \Gamma S^t$ respectively.

An equivalent, more physical, definition
can be given by the action of the squeezing operator ${\mathscr
S}(r,\phi)=\exp[\frac{r}{2}(\mbox{e}^{-i2 \phi} a^2-\mbox{e}^{i2
\phi} (a^\dagger)^2)]$ and the  displacement operator $\mathscr
D(u)$ defined above, on a thermal state
$\rho_{\beta}=(1-\ex{-\beta}) \sum_n \ex{-\beta n} \ketbrad{n}$,
where the Fock states $\ket n$ satisfy  $a^\dagger a
\ket{n}=n\ket{n}$:
\begin{equation}
\rho(\beta,\xi,r,\phi)=\mathscr D(\xi)^\dagger {\mathscr S}(r,\phi)
^\dagger\rho_{\beta}{\mathscr S}(r,\phi) \mathscr D(\xi)  .
\end{equation}
The covariance matrix of a thermal state is simply
$\Gamma_{\beta}=\gamma_{\beta} \id$, with $\gamma_{\beta}^{-1}=\tanh
(\beta/2)$. The squeezing operator $\mathscr S(r,\phi)$ induces  the
symplectic transformation  $S_{r,\phi}={O}_{\phi} D_{r} O_{\phi}^t
$, where
\begin{equation}
D_{r}=\left(\begin{array}{cc}\ex{r} & 0 \\0 &\ex{-r}
\end{array}\right), \quad O_{\phi}=\left(\begin{array}{cc}\cos\phi &
\sin\phi \\-\sin\phi &\cos\phi \end{array}\right),
\end{equation}
and the latter corresponds to a rotation in phase-space, i.e. to the
unitary operation $\mathscr O(\phi)=\exp[i\phi\,  a^\dagger a]$. One
thus finds that the covariance matrix can be written as
$\Gamma=\gamma_{\beta}S_{r,\phi}S_{r,\phi}^t$.

In order to calculate the Chernoff bound it is sufficient to realize
that any power $\rho^s$ of any Gaussian state $\rho$ is also a
Gaussian (unnormalized) state with a rescaled temperature:
\begin{eqnarray}
\rho(\beta,\xi,r,\phi)^s\!\!\!&=&\!\! \mathscr D(\xi)^\dagger \mathscr S(r,\phi) ^\dagger\rho_{\beta}^s\;
\mathscr S(r,\phi) \mathscr D(\xi)\nonumber \\
\!\!&=&\!\!N_{\beta,s} \mathscr D(\xi)^\dagger \mathscr S(r,\phi) ^\dagger\rho_{s \beta}
 \mathscr S(r,\phi) \mathscr D(\xi) \nonumber\\
\!\!&=&\!\!  N_{\beta,s}\; \rho(s \beta,\xi,r,\phi),
\end{eqnarray}
where we have used the relation
\begin{equation}
\rho_{\beta}^s=(1-\ex{-\beta})^s \sum_{n}\ex{-s\beta
n}\ketbrad{n}=N_{\beta ,s}\,\rho_{s \beta},
\end{equation}
with $N_{\beta,s}=(1-\ex{-\beta})^s/(1-\ex{-\beta s})$. Recall now
that given any two gaussian states $\rho_{A}$  and $\rho_{B}$, one
can write the inner product $\tr \rho_{A}\rho_{B}$ in terms of their
displacement vectors and covariance matrices as:
\begin{equation}
\tr(\rho_{A}\rho_{B})=2\left[\det(\Gamma_{A}+\Gamma_{B})\right]^{-\frac{1}{2}}\ex{-\delta^t
(\Gamma_{A}+\Gamma_{B})^{-1} \delta} , \label{inner}
\end{equation}
where $\delta=\xi_{A}-\xi_{B}$.
Using this equation we find that the quantum Chernoff
bound~(\ref{chernoff-quantum}) is $Q=\min_{s}Q_{s}$ with
\begin{eqnarray}
\kern-1.4em Q_{s}\!\!\!&=&\!\!\!\tr(\rho_{0}^s\rho_{1}^{1-s})\nonumber\\
\!\!\!&= &\!\!\!2N_{\beta_{0},s}N_{\beta_{1},1-s}
[\det(\tilde\Gamma_{0}+ \tilde\Gamma_{1})]^{-\frac{1}{2}}\ex{\delta^t (\tilde\Gamma_{0}
+\tilde\Gamma_{1})^{-1} \delta},
\end{eqnarray}
where $\tilde\Gamma_{i}=\gamma_{s\beta_{i}} S_{r_{i},\phi_{i}}S_{r_{i},\phi_{i}}^t$,
$i=0,1$, 
and $\delta=\xi_{0}-\xi_{1}$. To simplify the notation we will
denote the covariance matrix of the Gaussian state with $\beta=0$ as
$A= S_{r,\phi}S_{r,\phi}^t$.

\subsection{States with equal covariance matrices}

If  two general Gaussian states $\rho_0$ and $\rho_{1}$ are
identical modulo a relative displacement $\delta$, i.e. $\rho_{1}=
\mathscr D(\delta)\rho_{0} \mathscr D(\delta)^\dagger$  we find that
\begin{equation}
Q_{s}=\ex{-\delta^t (\tilde\Gamma_{1}+\tilde\Gamma_{2})^{-1}
\delta}=\ex{-(\gamma_{s\beta}+\gamma_{(1-s)\beta })^{-1}\delta^t
A^{-1}\delta } , \label{Qseqc}
\end{equation}
where in the first equality we used the fact that the factor
multiplying the exponential in \eqref{Qseqc} must be equal to one,
since it is independent of~$\delta$ and for~$\delta=0$ one must have
$\rho_{0}=\rho_{1}$, which implies that $Q_{s}=1$. That is,
\begin{eqnarray}
2 N_{\beta, s}N_{\beta ,1-s}&=&[\det(\gamma_{s\beta} A+\gamma_{ (1-s) \beta} A)]^{\frac{1}{2}}=\nonumber\\
&=& \gamma_{s\beta} +\gamma_{(1-s)\beta }, \label{nb}
\end{eqnarray}
where we have used that symplectic transformations have unit
determinant, i.e.,~$\det A=\det(SS^t)=1$.
One readily sees that $Q_s$, Eq.~\eqref{Qseqc}, attains its minimum
at~$s^*=1/2$, hence we find that  in this case the Chernoff measure
is:
\begin{eqnarray}
Q\!\!\!&=&\!\!\min_{s}Q_{s}=\exp\left({-\frac{1}{2\gamma_{\beta/2}}\delta^t A^{-1}\delta }\right)\\
\!\!\!&=&\!\! \exp\left(-\frac{1}{2}\delta^t O_{\phi} D_{2r}^{-1} O_{\phi}^t\delta \tanh{\beta\over4}\right)\nonumber\\
\!\!\!&=&\!\!\exp\left[-\frac{|\delta|^2}{2}
(\ex{-2r}\cos^2\theta+\ex{2r}\sin^2\theta)\tanh{\beta\over4}\right]
,\nonumber
\end{eqnarray}
where~$\theta$ is the relative angle between the squeezing axis and
the displacement vector, i.e., if $\delta=O_{\varphi}(|\delta|,0)^t$
then~$\theta=\varphi-\phi$.

\subsection{States with the same temperature}

We can generalize the previous result to states that have the same
spectra, i.e., the same temperature ($\beta_{0}=\beta_{1}=\beta$).
In this case we can use~\eqref{nb} to find
\begin{eqnarray}
Q_{s}\!\!&=&\!\!(\gamma_{s\beta} +\gamma_{(1-s)\beta})\det[\gamma_{s\beta} A_{0}
+ \gamma_{(1-s)\beta } A_{1}]^{-\frac{1}{2}}\nonumber\\
\!\!& \times&\!\! \exp\left[{\delta^t (\gamma_{s\beta} A_{0}+
\gamma_{(1-s)\beta } A_{1})^{-1} \delta} \right] .
\end{eqnarray}
The determinant can be explicitly written in a compact form as
\begin{eqnarray}
\det[\gamma_{s\beta} \id+ \gamma_{(1-s)\beta} {\mathscr A}]
\!\!&=&\!\!\gamma_{s\beta}^2+\gamma_{ (1-s)\beta}^2\nonumber\\
\!\!&+&\!\!2  \gamma_{s\beta}\gamma_{(1-s)\beta}\cosh( 2R),
\end{eqnarray}
where we have defined
\begin{equation}
{\mathscr
A}=S_{r_{0},\phi_{0}}^{-1}S_{r_{1},\phi_{1}}(S_{r_{0},\phi_{0}}^{-1}S_{r_{1},\phi_{1}})^t\equiv
S_{R,\Phi}{S_{R,\Phi}}^t,
\end{equation}
with
\begin{eqnarray}
\cosh 2R&=&\cos^2(\phi_{0}-\phi_{1})\cosh[2 (r_{0}-r_{1})]\nonumber\\
&+&\sin^2(\phi_{0}-\phi_{1})\cosh[2 (r_{0}+r_{1})].
\end{eqnarray}

With this generality $s^*$, the optimal value of $s$, is a
complicated function of the states' parameters \footnote{In contrast
to the claims in Exercise 3.9 page 77 of~\cite{hayashi}, it is not
generally the case that for states with equal spectra the minimum of~$Q_{s}$ is reached for~$s^*=1/2$.}.
In the case of~$\delta=0$, i.e., states with no relative
displacement and the same temperature, the minimization over~$s$ can
be done analytically, and one finds~$s^*=1/2$.  The quantum Chernoff
measure becomes:
\begin{eqnarray}
&&\kern-1em Q=\frac{1}{\cosh R}\\
&&\kern-1em=\left[ \cosh^2(r_{0}-r_{1})+\sin^2(\phi_{0}-\phi_{1}) \sinh
2r_{0}\sinh 2r_{1}\right]^{-1/2}. \nonumber 
\label{isospectral}
\end{eqnarray}
Notice that this expression is independent of the temperature (or
purity) of the states. That is, the distinguishability of two
arbitrary Gaussian states with no relative displacement and equal
temperature is independent of the degree of mixedness of the states.

\subsection{Chernoff metric for Gaussian states}

Following the definition~\eqref{ds2a} and using the previous
results we find that Chernoff metric is
\begin{eqnarray}
ds^2_{\rm QC}&=&\frac{d\beta^2}{32\sinh^2\frac{\beta}{2}}+\frac{dr^2+d\phi^2 \sinh^2 2r}{2}  \nonumber\\
& &+\frac{  \ex{-2r}dq_{\phi}^2+\ex{2r}
dp_{\phi}^2}{2}\tanh\frac{\beta}{4},
\end{eqnarray}
where we have defined the rotated displacement variables
$(q_{\phi},p_{\phi})=(q,p) O_{\phi}$ and we have used that for
infinitesimal changes $s^*=1/2$. We find again that the metric is
independent of the temperature under variations of the squeezing
parameters~$r$ and~$\phi$.

The (unnormalized) quantum Jeffreys prior can be obtained from the
metric tensor:
\begin{equation}
{\cal P}_{\rm QC}(\rho)\propto\sqrt{|\det \gb
|}=\frac{1}{16\sqrt{2}}\frac{\tanh\beta/4}{\sinh\beta/2} \sinh 2r .
\end{equation}

The metric induced by the local measure on the set of mixed states
is given by one-half the Bures metric~\footnote{There seems to be a
typo in \cite{kwek_quantum_1999} in the contribution of small
displacements of~Eq.~(13).}
\begin{eqnarray}
ds^2_{\rm CC}&=&\frac{d\beta^2}{32\sinh^2\frac{\beta}{2}}+
\frac{ \ex{-2r}dq_{\phi}^2+\ex{2r} dp_{\phi}^2}{4}\tanh\frac{\beta}{2} \nonumber\\
&+&\frac{dr^2 +d\phi^2\sinh^22r}{4} (1+{\rm sech}\beta) .
\end{eqnarray}
We note that, $ds^2_{\rm CC}\to \half ds^2_{\rm QC}$ as $\rho$
approaches the set of pure states ($\beta\rightarrow \infty$)  along
the lines~$d\beta=0$, in agreement with the general statement at the
end of~Sec.~\ref{sub-CC:BM}. In the limit of very mixed states
($\beta\approx 0$) the quantum Chernoff and local metric coincide up
to first order in~$\beta$. In this limit of high temperatures
($\beta\approx 0$,  highly mixed states) the quantum Chernoff metric
and Jeffreys prior agree with those derived from Bures distance
(modulo the omnipresent factor~$1/2$). In particular this implies
that the analysis in~\cite{slater_high-temperature_2000} of the
Bures volume element  in this high temperature regime also applies
here.

\section{Summary and conclusions}\label{C}

We have analyzed quantum state discrimination (symmetric hypothesis
testing) and the classical and quantum Chernoff bound focussing on
the  link between them and the concept of measures (distances) and
metrics on the space of quantum states. More precisely, we have been
concerned with defining measures and metrics that have a clear
operational meaning, so that they can as a matter of principle be
obtained from experiments. The error probability in state
discrimination, or rather its asymptotic rate exponent (error
exponent), has been shown to provide the natural  link. Thus, the
concept of {\em distinguishability measure} has emerged and has been
analyzed in depth throughout the central part of this work. Before
doing so, we have reviewed the methods and the main results of
classical and quantum hypothesis testing in the first three sections
of the paper. Qubit and Gaussian states have provided two excellent,
very relevant  examples to illustrate our results in the last
sections.

Our main points and results are summarized as follows: The quantum
Chernoff bound gives an upper bound to the error probability in
state discrimination.  When the unknown state (which we are asked to
identify as either one or the other of two known states) is a tensor
product, corresponding to many identical copies, the quantum
Chernoff information (which is essentially the log of the quantum
Chernoff bound) gives the error exponent of the optimal
discrimination protocol. We propose this quantity as a
distinguishability measure for general mixed states. We show that
the quantum Chernoff  measure is not attainable  by protocols that
use local fixed measurements (those for which the same measurement
is performed on each of the individual copies). Given the practical
relevance of these types of protocols (they can be realized with
current technology), we define a local distinguishability measure as
the error exponent of the best such protocol and present its main
features. We derive the metrics induced by these measures and their
corresponding volume elements. The latter provide a means to define
operational prior probability distributions of density matrices.  We
derive them for general matrices of arbitrary dimension.

Examples of all the above are given in the last part of~the paper.
For qubit and Gaussian states, we give explicit formulas for the
distinguishability measures and their corresponding metrics and
volume elements. We give a geometrical picture of the space of qubit
states based on those metrics. This space can be viewed as a
spherical cap, similar to Uhlmann hemisphere, with the pure states
sitting on the rim.  These examples also illustrate the fact that the
quantum Chernoff measure, besides being the most natural distance
between general states, is conveniently easy to compute relative to
other distances, such as the widely used fidelity.

\vspace{-.30em}

\section{Acknowledgments}

\vspace{-.30em}

We are grateful to Montserrat Casas, Juli C\'espedes,  Alex Monr\`{a}s,
Sandu Popescu and Andreas Winter for discussions. We are specially
grateful to Koenraad Audenaert and Frank Verstraete for their
collaboration at the early stages of this work.  We acknowledge
financial support from the Spanish MEC, through the Ram\'on y Cajal
program (JC), the travel grant PR2007-0204 (EB), contracts
FIS2005-01369, FIS2004-05639 (AC) and project QOIT (Consolider-Ingenio 2010), from the Generalitat de Catalunya, contract CIRIT SGR-00185 and from EU QAP project (AC).

\bibliography{chernoff4}

\end{document}